\documentclass[fleqn,usenatbib]{mnras}
\usepackage{newtxtext}
\usepackage{newtxmath}
\usepackage[T1]{fontenc}
\usepackage{ae,aecompl}
\usepackage{graphicx}	% Including figure files
\usepackage{amsmath}	% Advanced maths commands
\usepackage{soul}
\usepackage{xcolor}

\def\eagle{{\sc eagle}}

\title[Gas turbulence in simulated discs]{The physical drivers of gas turbulence in simulated disc galaxies}

\author[Esteban Jim\'{e}nez et al.]{
Esteban Jim\'{e}nez$^{1,3}$\thanks{E-mail: esteban.jimenez@icrar.org},
Claudia del P. Lagos$^{1,3,4}$,
Aaron D. Ludlow$^{1,3}$ and
Emily Wisnioski$^{2,3}$
\\
$^{1}$International Centre for Radio Astronomy Research (ICRAR), M468, University of Western Australia, 35 Stirling Hwy, Crawley, WA 6009, Australia\\
$^{2}$Research School of Astronomy and Astrophysics, Australian National University, Cotter Road, Westin Creek, ACT 2611, Australia\\
$^{3}$ARC Centre of Excellence for All Sky Astrophysics in 3 Dimensions (ASTRO 3D)\\
$^{4}$Cosmic Dawn Center (DAWN)\\
}

\date{Accepted XXX. Received YYY; in original form ZZZ}

\pubyear{2022}

\begin{document}
\label{firstpage}
\pagerange{\pageref{firstpage}--\pageref{lastpage}}
\maketitle

% Abstract of the paper
\begin{abstract}

We use the \eagle\ cosmological simulations to study the evolution of the vertical velocity dispersion of cold gas, $\sigma_{z}$, in central disc galaxies and its connection to stellar feedback, gravitational instabilities, cosmological gas accretion and galaxy mergers. To isolate the impact of feedback, we analyse runs that turn off stellar and (or) AGN feedback in addition to a run that includes both. The evolution of $\sigma_z$ and its dependence on stellar mass and star formation rate in \eagle\ are in good agreement with observations. Galaxies hosted by haloes of similar virial mass, $\rm M_{200}$, have similar $\sigma_z$ values even in runs where feedback is absent. The prevalence of local instabilities in discs is uncorrelated with $\sigma_z$ at low redshift and becomes only weakly correlated at high redshifts and in galaxies hosted by massive haloes. $\sigma_z$ correlates most strongly with the specific gas accretion rate onto the disc as well as with the degree of misalignment between the inflowing gas and the disc's rotation axis. These correlations are significant across all redshifts and halo masses, with misaligned accretion being the primary driver of high gas turbulence at redshifts $z \lesssim 1$ and for halo masses $\rm M_{200} \lesssim 10^{11.5} M_{\odot}$. Galaxy mergers increase $\sigma_z$, but because they are rare in our sample, they play only a minor role in its evolution. Our results suggest that the turbulence of cold gas in \eagle{} discs results from a complex interplay of different physical processes whose relative importance depends on halo mass and redshift.

\end{abstract}

\begin{keywords}
methods: numerical -- galaxies: evolution  -- galaxies: ISM -- galaxies: kinematics and dynamics 
\end{keywords}

%%%%%%%%%%%%%%%%% BODY OF PAPER %%%%%%%%%%%%%%%%%%

\section{Introduction} 
The assembly of galactic discs is regulated by gas accretion, star formation, feedback from stars and active galactic nuclei (AGN), and galaxy mergers. The existence of a tight main sequence (MS) between the star formation rates (SFR) and stellar masses ($\rm M_\star$) of galaxies suggests that discs assembled in quasi-equilibrium between these processes (\citealt{Bouche10}; \citealt{Dave12}; \citealt{Lilly13}; \citealt{Dekel14}; \citealt{Forbes14b}; \citealt{Rodriguez-Puebla16}; \citealt{Tacchella20}; \citealt{Wang22}; although see \citealt{Kelson14}). However, deciphering the relative importance of each process requires a precise understanding of the underlying physical conditions of discs across time. In this regard, valuable knowledge into the physical properties of discs can be gained through observations of the internal kinematics within galaxies \citep[see][for a review]{Glazebrook13}.

Insights on the kinematic properties of galaxies have been derived from long-slit spectroscopy \citep[e.g.][]{Davis03, Kassin07, Simons16, Kriek15} and Integral Field Spectroscopy (IFS) surveys. The latter in particular,  have allowed the study of kinematic properties, gas content and SFR at sub-galactic scales. For instance, ionised gas emission obtained by several IFS surveys has been used to study the kinematic evolution of star-forming galaxies (SFGs) at redshifts $z = 1-3$ \citep[e.g.][]{FS09, Wisnioski15, Stott16, Turner17}. One important finding of these studies is that by $z\approx 2$, many disc galaxies are already dominated by ordered rotation comparable to that of their low-redshift counterparts \citep[e.g.][]{FS06, FS09, Cresci09, Law09, Wisnioski15}. Rest-frame UV images of these structures indicate that they also contain several star-forming clumps \citep[e.g.][]{Elmegreen07, Fisher17}.

The “turbulent” nature of the interstellar medium (ISM) of high redshift galaxies can be observed in the velocity dispersion of their ionised gas. Data compilations from long-slit and IFS surveys suggest that ionised velocity dispersion increases approximately linearly with increasing redshift from $z=0$ to $z=3$ (\citealt{Kassin12}; \citealt{Wisnioski15}; \citealt{Simons17}; \citealt{Ubler19}; although see \citealt{DiTeodoro16}). Collectively, the observational data suggests that $z\approx 2$ galaxies exhibit ionised gas velocity dispersion $2-5$ times higher than their local counterparts. A similar evolution (although shifted to systematically lower values of velocity dispersion) is inferred from atomic hydrogen (HI) \citep{Dib06, Mogotsi16} or carbon monoxide (CO) \citep[e.g.][]{Leroy09, Swinbank11, Tacconi13}  emission, indicating that the cold-phase of ISM also becomes more turbulent at higher redshifts \citep[see also][]{Ubler18, Girard21}. 

%Bases on the {\sf KMOS$^{\rm 3D}$} data \citep{Wisnioski19}, \citet{Ubler19} found that the ionised velocity dispersion increases approximately linearly with increasing redshift. When combining results from various long-slit and IFS studies (see their fig. 6), the authors found a consistent dependency on redshift. Collectively, the observational data suggests that high-redshift galaxies exhibit ionised gas velocity dispersions $2-5$ times higher than their local counterparts}. 

%The “turbulent” nature of the interstellar medium (ISM) of high redshift galaxies is observed in the velocity dispersion of their ionised gas, which is a factor of 2−5 higher than in local SFGs. Using the KMOS3D sample, \citet{Ubler19} showed that the ionised velocity dispersion increases approximately linearly with increasing redshift. A similar evolution (although shifted to systematically lower values of velocity dispersion) is inferred from HI \citep{Dib06, Mogotsi16} or CO \citep[e.g.][]{Leroy09, Swinbank11, Tacconi13} emission, indicating that the cold-phase of the ISM also becomes more turbulent at higher redshifts \citep[see also][]{Ubler18, Girard21}. 

The dissipational nature of the ISM of galaxies suggests that turbulence should decay on a time scale comparable to the dynamical time ($\sim$ 100 Myr) of the galaxy disc with a \citet{Toomre64} stability parameter $Q\approx 1$ \citep{Krumholz18}. In contrast, turbulence should decay on shorter timescales ($\sim$ 10 Myr) in giant clumps. Hence, a continuous injection of energy into the ISM is necessary to sustain high levels of turbulence for several Gyr \citep[e.g][]{MacLow98, Stone98}. A number of energy sources have been proposed to explain the evolution of gas velocity dispersion in discs. The main contributors can be grouped in three categories: (i) the injection of feedback energy from massive stars and AGN, (ii) dynamical effects driven by local gravitational instabilities and (iii) cosmological accretion via cold flows.  

Star formation feedback can drive turbulence via the injection of kinetic and thermal energy from stellar winds, radiation pressure and supernovae (SNe) explosions, where the latter is thought to be the dominant effect \citep[e.g][]{MacLow04, Ostriker11}. The observed correlation between the velocity dispersion and global SFRs of gaseous discs supports this hypothesis \citep[e.g][]{Dib06, Green10, Green14, Johnson18, Law22}. There is an ongoing debate about whether feedback has a local or global effect on gas turbulence; some studies have found a weak correlation between the SFR surface density ($\Sigma_{\rm SFR }$) and gas velocity dispersion \citep[e.g.][]{Genzel11, Zhou17, Ubler19}, while others claim that there is a significant correlation \citep[e.g.][]{Lehnert13, Varidel20}.

Gravitational disturbances due to local instabilities may also act to increase the velocity dispersion \citep[e.g.][]{Aumer10}. Various studies have shown that local disc instabilities induce clump formation that generate radial flows within the disc enabling gas in the disc's outskirts to release gravitational potential energy and fall towards the disc's centre \citep[e.g.][]{Ceverino10, Krumholz10}. This mechanism is expected to be more important at high redshifts when discs had higher gas fractions and were therefore more vulnerable to forming local instabilities. Indeed, it appears that most high redshift disc galaxies contain a large number of massive, dense clumps in which most of their star formation occurs \citep[e.g.][]{Genzel11}.  

In addition to these internal drivers of turbulence, various external sources have also been considered, among them are cosmological gas accretion onto discs and galaxy interactions, such as major or minor mergers. Gas accretion via cold flows from the cosmic web is needed to replenish discs with fresh gas but, prior to settling into a rotationally supported structure, can contribute to disc turbulence \citep[e.g][]{Forbes23}. The relationship between turbulence and gas accretion is complex and depends on the smoothness of the cold streams \citep[e.g.][]{Mandelker18, Mandelker20}, on the density contrast between the stream and the disc \citep{Klessen10}, and as we show in Section~\ref{sec: PhysicalDrivers}, on the orientation of the accreting material with respect to the disc plane. Finally, galaxy mergers may also play a role by changing the dynamical and morphological structure of discs \citep[e.g.][]{DiMatteo11, Lagos18}. 

If discs remain in quasi-equilibrium as they grow, the evolution of their SFR and gas content can be related to the evolution of their gas velocity dispersion. Under this framework, \citet[][hereafter K18]{Krumholz18} developed an analytic model for the evolution of galactic discs that accounts for stellar feedback and radial gas transport originated from disc instabilities as the two main sources of gas turbulence. In their model, both sources of turbulence are needed to explain the observed evolution of the velocity dispersion of gaseous discs. \citet{Ginzburg22} extended K18's model by incorporating an analytical prescription to inject turbulence generated by cosmological gas accretion. Depending on redshift and the mass of the disc's parent dark matter (DM) halo, they found that gas accretion can indeed be an important driver of turbulence, especially at high redshift.  

To validate these analytic prescriptions across a broad range of environments, masses and redshifts it is necessary to have access to statistically representative samples of disc galaxies. Due to improvements in resolution and simulation volume, and to the implementation of sophisticated sub-resolution prescriptions for unresolved physics, the latest generation of cosmological simulations are useful tools for studying the evolving dynamics of gaseous discs. For example, \citet{Pillepich19} used the TNG50 simulation to study the evolution of gas velocity dispersion in SFGs, finding a similar evolution with redshift as that inferred from IFS surveys. However, the connection between turbulence and the various physical drivers mentioned above has not been studied using hydrodynamical simulations of cosmologically representative volumes.

In this paper, we use the \eagle{} simulations \citep{Schaye15, Crain2015} to study the evolution of gas turbulence and how it is related to the assembly of gaseous discs. We focus on two main questions: (i) what is the relation between gas velocity dispersion and the various potential drivers of turbulence, such as the SFR, mergers, or gas accretion?; and (ii) do these relations depend on halo mass and/or redshift? \eagle{} is a suitable tool to carry out this analysis as it reproduces well the main sequence of star formation from $z=0$ to $z\approx 5$ \citep{Furlong15, DSilva23}, the SFR and stellar mass functions from $z=0$ to $z\approx 4$ \citep{Furlong15, Katsianis17}, the abundance of molecular gas as a function of cosmic time \citep{Lagos15} and its relation to SFR and $\rm M_\star$ \citep{Lagos16} at $z=0-4$, and stellar kinematics properties of galaxies \citep[e.g.][]{Ludlow2017,Lagos17, Lagos18b, Swinbank17, WaloMartin20}.

The outline of this paper is as follows. In Section~\ref{Methods} we introduce the \eagle\ simulation, define our galaxy sample and explain how we measure the velocity dispersion of gaseous discs. In Section~\ref{SecEvsigma} we compare the gas velocity dispersions measured for our sample of \eagle{} galaxies with those inferred from various observational data sets, and also present scaling relations between the gas velocity dispersion and various other galaxy properties. In Section~\ref{sec: PhysicalDrivers} we present our main results, providing an interpretation and discussion in Section~\ref{sec: Discussion};  Section~\ref{sec: Conclusions} contains our conclusions. 

% ==============================================================
%       SECTION 2: METHODS
% =============================================================

%note that many surveys assume that sigma is uniform across the disc

\section{Methods} \label{Methods}

\subsection{The \eagle\ simulations}

\eagle\ is a suite of cosmological, smoothed particle hydrodynamical (SPH) simulations that follow the assembly of DM haloes and the formation of galaxies within them \citep{Schaye15, Crain2015}. The simulations were carried out using a modified version of the $N$-body SPH code GADGET-3 \citep{Springel05, Springel08}, employing cosmological parameters obtained by the \citet{Planck14}. The various \eagle\ runs adopt a range of cosmological box sizes, mass and force resolutions, and subgrid physics models, and are comprised of 28 outputs (snapshots) spanning $z=20$ to $z=0$. Table~\ref{tab: eagle-runs} provides the relevant details of the \eagle{} runs used in this paper. We also use the \eagle\ snipshots, a set of $200$ lean outputs spanning the same redshift range, to assess whether our results are sensitive to the time cadence of the standard \eagle\ snapshots.

\begin{table}
    \centering
    \begin{tabular}{l|ccc}
        \hline 
        Parameter & {\sc NoFb-L25} & {\sc NoAGN-L50} & {\sc Ref-L100}\\
        \hline
        $L$ [cMpc] & 25 & 50 & 100\\
        $N_{\rm part}$ & $2\times 376^3$  & $2\times 752^3$ & $2\times 1504^3$\\
        $m_{\rm gas}$ $\rm [M_{\odot}]$ &  $1.81\times 10^6$ &  $1.81\times 10^6$ & $1.81\times 10^6$\\
        $m_{\rm DM}$ $\rm [M_{\odot}]$ & $9.70\times 10^6$ & $9.70\times 10^6$ & $9.70\times 10^6$ \\
       $\epsilon_{\rm com}$ $\rm [ckpc]$& 2.66 & 2.66 & 2.66\\
       $\epsilon_{\rm prop}$ $\rm [pkpc]$& 0.70 & 0.70 & 0.70\\
       \hline
    \end{tabular}
    \caption{\eagle\ simulation runs used in this paper. Rows from top to bottom show: the simulation name suffix; the box comoving size; the initial number of DM and gas particles; initial mass of gas particles; DM particle mass; comoving, Plummer-equivalent gravitational softening length at $z\geq 2.8$; and maximum proper softening length at $z<2.8$. Here, cMpc, ckpc and pkpc refer to comoving megaparsec and kiloparsec, and proper kiloparsec, respectively.}
    \label{tab: eagle-runs}
\end{table}

DM haloes were identified in each snapshot using a Friends-of-Friends (FOF) algorithm \citep{Davis85} with a linking length of $0.2$ times the mean (Lagrangian) DM inter-particle separation. SUBFIND \citep{Springel01} was then run on the FOF haloes to identify gravitationally bound DM subhaloes, which are the potential hosts of galaxies. The baryonic content of these subhaloes was determined by associating each baryonic particle (gas or stellar) to its nearest DM particle, provided the latter is bound to a SUBFIND subhalo. The mass of each FOF halo is dominated by a central subhalo; the galaxy it contains (if any) is defined as the central galaxy. Each FOF halo also has a population of lower-mass satellite subhaloes, which are the potential hosts of satellite galaxies. Our analysis focuses exclusively on central galaxies.  

For each central galaxy and its DM halo, SUBFIND determines a number of relevant physical properties including its virial mass and radius, $\rm M_{200}$ and $r_{200}$ respectively, and the stellar and gas mass of its central galaxy. In what follows we define $\rm M_{200}$ as the mass contained within a sphere of radius $r_{200}$ that encloses an average density equal to $200$ times the critical density of the universe ($\rho_{\rm crit}=3 H^2/8 \pi G$, where $H$ is the Hubble-Lema$\hat{i}$tre constant and $G$ is the gravitational constant). The virial velocity, $V_{\rm 200}=\sqrt{G {\rm M_{200}}/r_{200}}$, corresponds to the circular velocity at $r_{200}$. Note that all particle types are included in the calculation of the virial quantities above. The stellar and gas masses of each central galaxy are calculated by summing the individual masses of each stellar or gaseous particle belonging to the central galaxy, excluding those that are bound to satellite subhaloes. 

The interplay of several physical processes governs the condensation of baryons in the central regions of DM haloes, their subsequent conversion into stellar particles and the build-up of galaxies. Specifically, \eagle\ models a variety of physical processes, such as radiative gas cooling and photoheating \citep{Wiersma09b}, star formation \citep{Schaye08}, stellar and chemical evolution \citep{Wiersma09}, stellar mass loss and feedback from supernovae \citep{DallaVecchia12}, the formation and growth of supermassive black holes (BHs), and AGN feedback \citep{Rosas-Guevara15}. Galaxy mergers and gas accretion are natural consequences of the simulation's cosmological initial conditions. 

The majority of our analysis is based on the $100$~cMpc Reference model of the \eagle{} simulation suite (hereafter the {\sc Ref-L100} run). Due to the finite resolution of cosmological simulations and our limited understanding of several physical processes relevant for galaxy evolution, such as stellar and AGN feedback, a set of sub-grid prescriptions for unresolved physical processes are usually employed. In general, the equations that implement these processes contain free parameters that must be calibrated so that simulation results reproduce important observations of the galaxy population. \eagle's sub-grid physics models were calibrated so that the simulation reproduced the observed local Universe stellar mass function, the galaxy size-$\rm M_\star$ relation, and BH mass-$\rm M_\star$ relations (see \citealt{Crain2015} for details). 

The \eagle\ suite also includes runs that adopt variations of the subgrid parameters employed for the Reference model. These runs were not required to match the observations mentioned above but can nonetheless be used to explore the effect of changing various subgrid parameters on the galaxy population (see \citealt{Crain2015} for a detailed discussion). We make use of two such runs: one in which feedback from stars {\em and} AGN was turned off (hereafter, {\sc NoFb-L25}), and another that includes stellar feedback but none from AGN (hereafter {\sc NoAGN-L50}). For both of these runs, the remaining subgrid parameter values were identical to those used in the {\sc Ref-L100} run. We use these runs to assess the impact of stellar and AGN feedback on the kinematics of gaseous disc galaxies. Note that the {\sc NoFb-L25} model was carried out in a $25$~cMpc simulation box, while the {\sc NoAGN-L50} was run in a $50$~cMpc cubed volume. 

We focus our analysis on the redshift range $0.1<z<4$, which overlaps with most kinematic measurements of gaseous discs from spectroscopic surveys (see Section~\ref{SecEvsigma}). Snapshots in the {\sc NoFb-L25} run are only available down to $z=0.1$; hence we adopt this as the lower redshift bound for most of our analysis, although we include $z=0$ results from the {\sc Ref-L100} and {\sc NoAGN-L50} runs for comparison with observations. We note that the quantities of interest for this work -- gas velocity dispersions, stellar and gas masses, etc -- evolve very little between $z=0.1$ and $z=0$.

Below, we describe the relevant aspects of \eagle's sub-grid models. 

% =====================================================
\subsection{Modelling star formation and feedback from stars and AGN}

Star formation in \eagle\ is implemented stochastically following the Kennicutt-Schmidt star formation relation reformulated as a pressure law \citep[see][for details]{Schaye08}. The pressure is determined using a polytropic equation of state, $P = P_{\rm eos}(\rho)$, normalised to a temperature floor ${\rm T}_{\rm eos}=8000\ {\rm K}$ at density $n_{\rm H}=0.1\ {\rm cm^{-3}}$.
Because \eagle\ does not resolve the cold phase of the ISM, the simulation triggers star formation stochastically in gas particles that exceed a metallicity-dependent gas density threshold, $n_{\rm H}^{*}(Z)$, 
\begin{equation} \label{eq: rho_crit}
n_{\rm H}^* = 10^{-1}\ {\rm cm^{-3}}\ \left(\frac{Z}{0.002}\right)^{-0.64}. 
\end{equation}

\noindent Here, $Z$ is the gas metallicity. The metallicity-dependence of equation~(\ref{eq: rho_crit}) approximately accounts for the increased cooling efficiency and enhanced shielding by dust grains that are expected in high-metallicity gas, which in practice lowers the density required for the transition from the warm to the cold phase of the ISM, and hence also lowers the density threshold for star formation \citep[e.g.][]{Richings14}. 

Stellar particles are considered as simple stellar populations characterised with a \citet{Chabrier03} initial mass function (IMF); stars with masses in the range $6-100\, \rm M_{\odot}$ are assumed to end their lives as core-collapse supernovae after $3\times 10^7\, \rm yr$ from their formation time.
\citet{Dalla-Vecchia12} introduced a stochastic approach to supernova feedback that allows the amount of thermal energy injected per supernova to be controlled in order to overcome numerical radiative losses due to poor resolution. The fraction of energy from a supernova that is injected into the neighbouring gas particles is governed by the feedback efficiency parameter, $f_{\rm th}$. A value of $f_{\rm th}=1$ indicates that all energy produced by supernovae is imparted to the gas, whereas a low value implies less efficient feedback. The exact value of $f_{\rm th}$ in \eagle\ depends on the local conditions of the ISM, such as the gas-phase metallicity and density. For the {\sc Ref-L100} and {\sc NoAGN-L50} runs, the mean values of $f_{\rm th}$ at $z=0.1$ are close to $1$, while for the {\sc NoFb-L25} run $f_{\rm th}$ is manually set to zero. 

The expectation value for the number of heated gas particles per supernova, $\left< N_{\rm heat}\right>$, is proportional to the feedback efficiency parameter, and is given by
\begin{equation} \label{eq: N-heated}
    \left< N_{\rm heat}\right> \approx 1.3 f_{\rm th} \left( \frac{\Delta T}{10^{7.5}{\rm K}}\right)^{-1},
\end{equation}
where $\Delta T=10^{7.5}\,\rm  K$ is the desired temperature increment of the heated gas particles. $\Delta T$ remains fixed and is chosen to be high in order to prevent gas overcooling. 

A BH seed of mass $\rm M_{BH}=1.5\times 10^5\, M_\odot$ is placed in the centre of every newly-formed DM halo whose FOF mass exceeds $\gtrsim 10^{10}\, {\rm M_\odot}$; they subsequently grow in mass by accreting neighbouring gas particles and by merging with other BH particles \citep[e.g][for details]{Springel05a}. As with stellar feedback, AGN feedback is implemented stochastically. In the case of AGN feedback, however, a higher temperature increment of $\Delta T=10^{8.5}\ \rm K$ is adopted \citep{Booth09}. This ensures AGN feedback remains efficient when the associated energy is injected into gas particles near the central BH, which typically have higher densities than those surrounding stellar particles and are therefore more prone to radiative losses. 

Gas accretion onto BHs is modelled following the Bondi-Hoyle accretion model \citep{Bondi-Hoyle44}. An efficiency parameter is introduced in the AGN feedback model which accounts for the amount of rest energy from the accreted gas that is injected into the surroundings. \citet{Rosas-Guevara15} introduced an additional dependence on the accretion-disc angular momentum which can suppress the accretion onto the supermassive BH and reduce the AGN feedback efficiency.

% ==================================================================
\subsection{Calculation of gas accretion rates} \label{subsec: CalcGasAcc}

{We link galaxy descendants and progenitors in adjacent snapshots using the galaxy merger trees available in the \eagle\ database \citep{McAlpine16,Qu17}. We use this information together with the particle data to compute gas accretion rates for each \eagle\ galaxy in our sample. We will use these accretion rates in Section~\ref{SecGasAcc} to study correlations between gas accretion and turbulence.}

We compute time-averaged accretion rates (i.e. inflows averaged over a finite time interval $\Delta \rm t$) using the spatial distribution of particles in adjacent snapshots, \{$j+1,\, j$\}, where $j+1$ refers to the lower redshift of the two snapshots; descendant galaxies therefore belong to snapshot $j+1$ and progenitors to snapshot $j$. In \eagle, the time between $j$ and $j+1$ ranges from $\approx 0.3\  {\rm Gyr}$ (for $z>2$) to $1\ {\rm Gyr}$ (for $z < 1$). As shown by \citet[][see also \citealt{Wright20}]{Mitchell20}, poor temporal resolution can affect estimates of accretion rates by not properly accounting for particles that had been accreted but subsequently lost on a timescale shorter than the time between adjacent snapshots. Using the \eagle\ snipshots\footnote{The time interval between snipshots ranges from $\approx 60$ to $\approx 130$~Myr.}, we verified that our accretion rates converge if averaged over a time interval that is of order or larger than the typical dynamical timescale of galaxies\footnote{The dynamical time is defined as $\tau_{\rm dyn}=\sqrt{3\pi/16G\bar{\rho}}$, where $\bar{\rho}$ is the total matter density enclosed by a $r=0.2\, r_{200}$ spherical aperture.}  (which is $\gtrsim 400$ Myr), in agreement with the conclusions of \citet{Mitchell20, Wright20}.

%\textcolor{red}{We used} the \eagle\ snipshots\footnote{\textcolor{red}{Note that the time interval between snipshots} ranges from $\approx 60$\textcolor{red}{\, to $\approx$}$130$ Myr.} \textcolor{red}{to verify that our} accretion rates are converged when averaged over a time interval that is \textcolor{red}{of order} or larger than the typical dynamical timescale of galaxies\footnote{\textcolor{red}{The dynamical time is defined as} $\tau_{\rm dyn}=\sqrt{3\pi/16G\bar{\rho}}$, where $\bar{\rho}$ is the {\textcolor{red}{total}} matter density enclosed \textcolor{red}{by} a $r=0.2\, r_{200}$ \textcolor{red}{spherical} aperture.}  (\textcolor{red}{which is} $\gtrsim 400$ Myr)\textcolor{red}{, in agreement with the conclusions of} \citet{Mitchell20, Wright20}}.

%However, accretion rates are well converged when averaged over longer time intervals (e.g. 1$\, {\rm Gyr}$), which is comparable to the temporal spacing between snapshots in \eagle\ at low redshifts. 

We calculate the total gas mass transferred from the circum-galactic medium (CGM) to the ISM between consecutive snapshots as follows. First, for each galaxy in snapshot $j+1$, we select all gas particles with temperature ${\rm T} \leq 10^4\ {\rm K}$ {\em or} a non-zero SFR (our definition for ``cold gas''; see justification below) contained in a sphere of radius $0.2\,r_{200}$\footnote{Note that we use the centre of potential and $r_{200}$ values of the progenitor galaxy in the snapshot $j$} (which we use as a boundary to delineate the disc and CGM). Then, we track their 3D positions to their main progenitor galaxy in snapshot $j$ and considered as accreted particles those whose radial separation from the progenitor's centre of potential (COP) exceeds $0.2\, r_{200}$ (note that we do not impose any condition on its temperature). We also account for gas particles that were converted into star particles in the disc during the interval $\Delta t$ by selecting star particles in the disc in snapshot $j+1$ that were CGM gas particles at snapshot $j$.
The net accretion rate, $\dot{\rm M}_{\rm acc}$, is defined as the sum of the mass of the accreted particles (stellar or gas) divided by the time interval $\Delta t$ between the two snapshots.

%We calculate the total gas mass transferred from the circum-galactic medium (CGM) to the ISM between consecutive snapshots as follows. For each galaxy in snapshot  $j+1$, we select all coldgas particles in the disc and track their 3D positions to their main progenitor galaxy in snapshot $j$. Particles whose radial separation from the progenitor’s centre of potential (COP) exceeds $0.2\,r_{200}$ (which we use as a boundary to delineate the disc and CGM) are considered accreted particles regardless of their temperature. We also account for gas particles that were converted into star particles in the disc during the interval $\Delta t$ by selecting star particles in the disc in snapshot $j+1$ that were CGM gas particles at snapshot $j$. The net accretion rate, $\dot{\rm M}_{\rm acc}$, is defined as the sum of the mass of the accreted particles (stellar or gas) divided by the time interval $\Delta t$ between the two snapshots.

 % ==================================================================
\subsection{Galaxy sample selection}
\label{sec: GalSamples}

\begin{figure}
	\centering
	\includegraphics[width=\columnwidth]{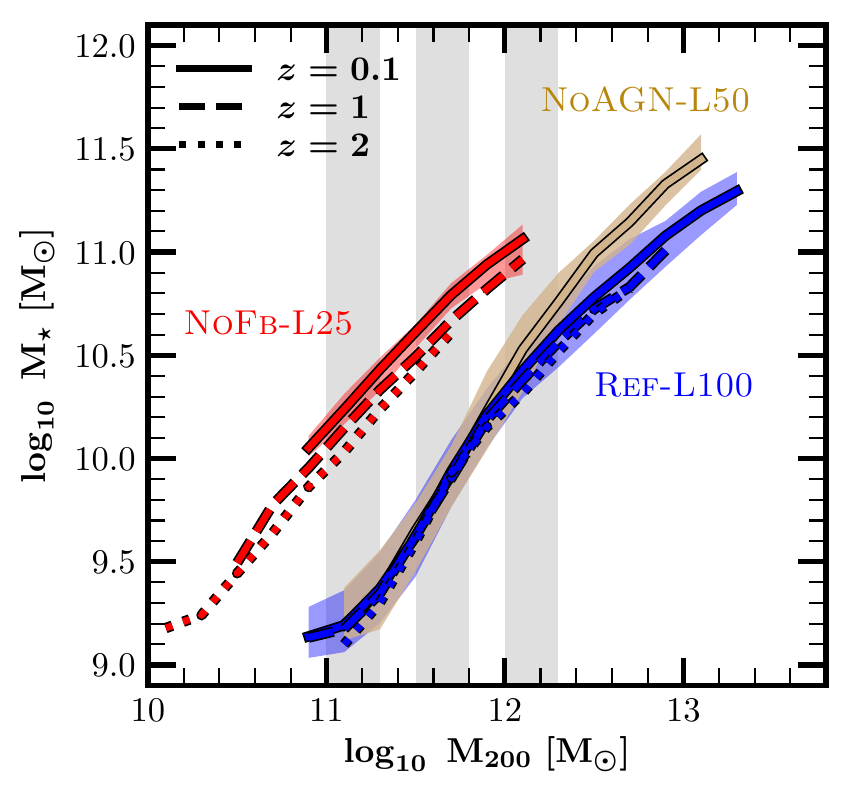}
	\caption{The $\rm M_{\star}-M_{200}$ relation in the {\sc Ref-L100} (blue), {\sc NoAGN-L50} (light brown) and {\sc NoFb-L25} (red) runs at $z=0.1$ (solid), $z=1$ (dashed) and $z=2$ (dotted). The relations are shown for galaxies that meet our selection criteria (see text for details). Lines show the medians in bins with $\geq 10$ galaxies while shaded regions (only shown for $z=0.1$) show the $16^{\rm th}$ and $84^{\rm th}$ percentiles. The vertical grey bands indicate the halo mass bins analysed throughout the paper. } 
	\label{fig: stellar-to-halo-mass rel}
\end{figure}

We focus our analysis on the cold gas content of central galaxies with $\rm M_{\star} \geq 10^9\,\rm  M_{\odot}$. We do not consider satellite galaxies in our analysis because they are more sensitive to external sources of turbulence, such as ram-pressure stripping and strong tidal interactions \citep[e.g.][]{Gunn72, Boselli06, Bahe15, Wright22}.
As mentioned above, we define the cold gas component of galaxies as the subset of gas particles that either have a temperature ${\rm T}\leq 10^4\, {\rm K}$ or ${\rm SFR}>0$\footnote{Selecting star-forming gas particles (which can potentially convert into star particles) implies selecting regions that are likely to be exposed to radiation from massive stars, which ionises the surrounding gas creating HII regions; our selection of ``cold gas'' particles therefore approximately traces all phases of the ISM.}. The first condition approximately selects the warm phase of the ISM which is primarily composed of atomic hydrogen; the latter condition targets the unresolved cold phase which is most likely to  contain the molecular hydrogen component. These criteria for identifying the cold gas phase of simulated galaxies are widely adopted in the literature \citep[e.g.][]{Wright21}.

We also focus our analysis on galaxies that harbour a prominent gaseous disc, in line with the majority of observational studies. To identify discs we use the kinematic indicator $\kappa_{\rm co}$, introduced in \citet{Correa17}, which quantifies the fraction of the disc's kinetic energy that is invested in co-rotation, i.e. 
\begin{equation} \label{eq: kappa_co}
    \kappa_{\rm co} = \frac{1}{K} \sum_{i, L_{z,i}>0} \frac{1}{2}m_{i} \left(\frac{L_{z,i}}{m_{i} R_{ i}}\right)^2,
\end{equation}

\noindent where $K=0.5\sum_{i} m_{i}\, v^2_{i}$ is the total kinetic energy of the particles, $R_{i}=(r_{i}^2-z_i^2)^{1/2}$ is the distance of the $i^{\rm th}$ particle to the galaxy's net angular momentum vector (which we align with the $z$-axis, in our analysis), and $L_{z,i}$ is the $z$-component of the particle's angular momentum. The sum extends over all particles of the relevant species that lie within a spherical aperture of radius $r=2\, r_{50}$ and also have a positive $L_{z}$ ($r_{50}$ is the three dimensional half-mass radius of the galaxy's gas component).

\citet{Correa17} applied equation~(\ref{eq: kappa_co}) to stellar particles in \eagle\ galaxies and found that $\kappa_{\rm co}\gtrsim 0.4$ approximately marks the transition between passive, spheroidal galaxies and star-forming disc galaxies. \citet{Thob19} showed that $\kappa_{\rm co}$ correlates with a number of other kinematics properties used to characterise simulated galaxies, such as the ratio of rotation to dispersion velocities, $v_{\rm rot}/\sigma_{\rm 1D}$, the stellar spin parameter, $\lambda_\star$, and the orbital circularity parameter $\xi=j_{z}/j_{\rm c}$. 

Unlike most previous studies, we follow \citet{Manuwal22} and use equation~(\ref{eq: kappa_co}) to characterise the morphologies of each galaxy's gas component. By visually inspecting edge-on projections of the cold gas distribution for a large sample of \eagle~ galaxies, we find that $\kappa_{\rm co}\geq 0.7$ singles-out thin gas discs. 
In \eagle, galaxies with $\kappa_{\rm co}\gtrsim 0.7$ typically have $v_{\rm rot}/\sigma_{\rm 1D}\gtrsim 2$.
We note that imposing a restriction on $\kappa_{\rm co}$ is necessary to select gaseous discs in {\sc Ref-L100}, but not for {\sc NoFb-L25}. Indeed, most galaxies in the latter run satisfy $\kappa_{\rm co}>0.9$, suggesting that, in the absence of feedback, cold gas is primarily concentrated in a thin (albeit compact) rotating disc. Applying this kinematic cut removes two main galaxy groups: (i) massive galaxies at low redshifts, which are generally passive, round-shaped, and have low gas content, and (ii) low-mass galaxies at high redshift, whose gaseous component has not yet settled into a rotating disc. 

To simplify the interpretation of our results -- and to allow us to isolate the impact of feedback, gravitational instabilities, and gas accretion on the vertical velocity dispersion of gaseous discs -- for most of our analysis we remove galaxies that have undergone recent mergers. We note, however, that this additional selection criterion does not affect our results, mainly because galaxy mergers are rare, affecting only 2 per cent of all galaxies in our sample at $z=0.1$. We consider the impact of merger on $\sigma_z$ explicitly in Section~\ref{SecGalMergers}.

Finally, we only consider galaxies whose dynamical properties -- particularly their vertical gas velocity dispersion, $\sigma_z$, the calculation of which we describe below -- are computed using at least $500$ cold gas particles. This reduces the effects of Poisson noise on estimates of the kinematic properties of galaxies that contain low numbers of cold gas particles. 

Imposing these criteria on all galaxies with $\rm M_\star\geq 10^9 \, M_{\odot}$ removes roughly $56$ per cent of galaxies at $z=0$, and $77$ per cent at $z=2$. Nevertheless, the remaining galaxies (of which there are $3277$ at $z=0$ and $1361$ at $z=2$) sample a wide range of galaxy properties. Indeed, we find that the star formation main sequence is well-sampled at all redshifts (for ${\rm M_{\star}} \ge 10^{9}\,\rm M_{\odot}$). 
%\textcolor{red}{Please double check that this section is still accurate. I moved the statement from S3.1 about removing merging galaxies to this section. There, you stated that mergers are removed because youo want to isolate the effects of SFR, accretion and instabilities, but none of those things are considered in Section 3, so it seems very out of place.}

Fig.~\ref{fig: stellar-to-halo-mass rel} shows the stellar-to-halo mass relation for all galaxies that meet our selection criteria. We show results at $z=0.1$, $z=1$ and $z=2$ and for the three \eagle{} runs listed in Table~\ref{tab: eagle-runs}. As expected, the {\sc NoAGN-L50} run produces higher stellar masses at fixed $\rm M_{200}$ than the {\sc Ref-L100} run at $\rm M_{200}\gtrsim 10^{12}\,\rm M_{\odot}$ due to the lack of AGN feedback. At fixed $\rm M_{\star}$, the galaxies in {\sc NoFb-L25} are biased towards low halo masses. The lack of massive haloes in this run is due to the small simulation volume, which is only 25 cMpc on side. Beyond volume, the clearest difference between the  {\sc NoFb-L25} and the other two runs is the much larger $\rm M_{\star}$ of galaxies at fixed $\rm M_{200}$, which is expected when feedback is absent and therefore unable to suppress gas accretion and star formation. This is important to keep in mind when comparing galaxies at fixed $\rm M_{\star}$ between runs. Fig.~\ref{fig: stellar-to-halo-mass rel} also highlights three halo mass bins (vertical grey bands) that we use throughout the paper when comparing galaxies across all three simulations. 

For the {\sc Ref-L100} run, there are $\gtrsim 300$ galaxies in all mass bins and at all redshifts, except at $z\approx 2$ where there are $\approx 150$ in the highest halo-mass bin. The number of galaxies in the {\sc NoAGN-L50} run varies between $50-150$ at low redshifts and at all halo masses, but drops to $\approx 20$ for the lowest halo mass bin at $z\approx 2$. Because of the small volume of the {\sc NoFB-L25} run, we performed most of our analysis using galaxies in the low-mass bin, which contains $\gtrsim 60$ galaxies at all redshifts. Note that in some figures, we stack results from multiple adjacent snapshots which increases the number of galaxies analysed by a factor of $\approx 2-3$.

% ===================================================
\subsection{Calculation of the ISM gas velocity dispersion} \label{sec: sigma_z calculation}

We compute the vertical ISM gas velocity dispersion, $\sigma_z$, which is the component perpendicular to the plane of the galaxy's cold gas disc. To do so, we calculate the relative velocities of the gas particles with respect to the centre of mass (COM) of their host galaxy: $\Delta v = |\vec{v} - \vec{v}_{\rm COM}|$. The COM velocity vector, $\vec{v}_{\rm COM}$, is determined using stars, gas and DM particles within a sphere of radius $0.2\, r_{200}$ centred in the COP of the halo, as defined by SUBFIND. We define the disc plane as the plane perpendicular to the total angular momentum vector of the cold gas and young stars ($< 100$ Myr old). The latter is computed using the particles enclosed in the cold gas 3D half-mass radius, $r_{50}$. We reset all particle positions and velocities with respect to the COM reference frame and align the $z$-axis with the angular momentum of the disc as defined above. We then select cold gas particles that lie within an appropriately-defined cylinder and compute a global mass-weighted velocity dispersion using
\begin{equation} \label{eq: sigmaz}
    \sigma_z = \sqrt{\frac{\sum_i m_i \left[v_{z,i}^2 + \sigma_{{\rm P},i}^2 \right]}{\sum_i m_i}},
\end{equation}
where $\sigma_{{\rm P},i}=\sqrt{{\rm P_i}/\rho_i}$ corresponds to the thermal component of the velocity dispersion, which depends on the density and pressure of the gas particle. %\footnote{The thermal motions are assumed to be isotropic, with $\rm \sigma_{{\rm P},i}/\sqrt{3}$ representing the one dimensional thermal contribution to the velocity dispersion.}.
Note that star-forming particles in the equation of state have artificially high pressure, exceeding that of cold gas. Therefore, for these particles, we set $\sigma_{{\rm P},i}= 8\ {\rm km\ s^{-1}}$ which is the sound speed of gas at a temperature of $8000\,{\rm K}$ (i.e. the temperature floor imposed in \eagle{}; see \citealt{Schaye15}). This $\sigma_{{\rm P},i}$ is characteristic of the warm neutral medium and an upper limit if the gas was allowed to cool down further. 

The dimensions of the cylinder over which equation~(\ref{eq: sigmaz}) is applied is determined as follows. First, the height of the cylinder, $z_{90}$, is chosen so that it encloses $90$ per cent of all cold gas particles in the galaxy. However, we impose a lower and higher limit to $z_{90}$ of $1\, {\rm kpc}$ and $6\,{\rm kpc}$, respectively. This ensures that the scale height is always larger than the gravitational softening length ($\approx 0.7\,{\rm kpc}$), and small enough to only encompass cold gas that is located close to the disc plane. We also note that, at high redshifts, a significant amount of cold gas is contained in clumps away from the disc plane; the upper limit of $6$~kpc excludes most of these clumps, which are not themselves part of the gas disc. Note that for all galaxies that meet our selection criteria. Overall, imposing these limits on $z_{90}$ excises $\approx 20\%$ of the cold gas mass for galaxies that satisfy our selection criteria but has no discernible impact on our conclusions.

With $z_{90}$ fixed, we compute $\sigma_z$ as a function of the distance, $R$, from the galaxy centre\footnote{In general, we use lower base $r$ to denote three dimensional radii and upper case $R$ to denote radii within the disc plane.}. For simplicity, we will refer to this function as a ``velocity dispersion profile'', $\sigma_z(R)$. Note that $\sigma_z(R)$ includes \emph{all} the cold particles enclosed by $R$, not those in a bin centered on $R$. It is important to note that this quantity should not be confused with the velocity dispersion profiles derived from IFS observations which use line-emission properties in each spaxel to infer the local gas turbulence as a function of radius. For each galaxy, we take the value of the velocity dispersion where the $\sigma_z(R)$ function converges to a roughly constant value. For most galaxies (see Fig.~\ref{fig: sigma_profiles} below) in the {\sc Ref-L100} run, this occurs at $R_{\rm flat}={\rm a\, few}\,R_{50}$, with $R_{50}$ being the cylindrical cold gas half-mass radius. In practice, we measure $\sigma_z$ at $R_{\rm flat}=3\,R_{50}$\footnote{This is different from $r_{\rm 50}$, which is the 3D cold gas half-mass radius}. Incidentally, when the cold gas component does not extend up to this radius, we measure $\sigma_z$ at the radius of the outermost gas particle from the galaxy's centre.

%This is what is usually done in observational studies and aims to exclude the central regions of galaxies where velocity dispersion measurements are susceptible to the effects of beam-smearing. 
%\footnote{This is different from $r_{\rm 50}$, which is the 3D cold gas half-mass radius. In general, we use lower base $r$ to denote three dimensional radii and upper case $R$ to denote radii within the disc plane.}. 

\begin{figure}
	\centering
	\includegraphics[width=\columnwidth]{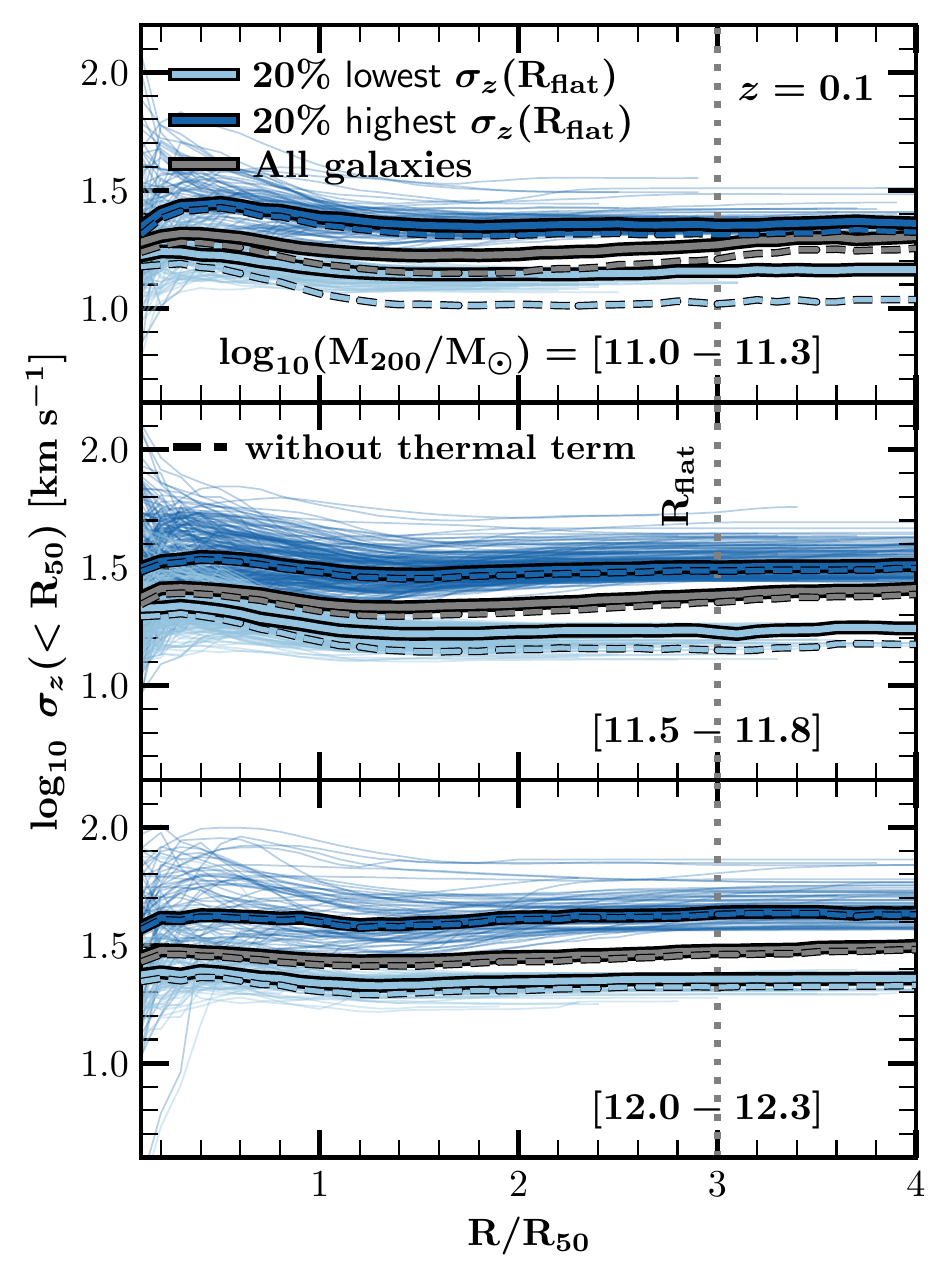}
	\caption{Vertical velocity dispersion profiles of cold gas for $z=0.1$ galaxies in the {\sc Ref-L100}. Solid (dashed) lines indicate the $\sigma_z$ profiles computed from equation~\ref{eq: sigmaz} with (without) the thermal component, $\sigma_{{\rm P},i}$. Different panels correspond to separate bins of $\rm M_{200}$, as labelled (the values are expressed in units of ${\rm log_{10}\, M_{\odot}}$). Radii have been normalised by $R_{50}$, cylindrical half-mass radius of cold gas. The vertical dotted lines indicate the radius $R_{\rm flat}=3\,R_{\rm 50}$ within which the profiles are evaluated. Thick grey curves show the median $\sigma_z(R)$ profiles for all disc galaxies in each mass bin; thick light (dark) blue lines correspond to the lower (upper) quintile of the $\sigma_z(R_{\rm flat})$ distribution in the same mass bins (thin lines of corresponding colour show individual profiles in these subsamples).} 
	\label{fig: sigma_profiles}
\end{figure}

\begin{figure}
	\centering
	\includegraphics[width=\columnwidth]{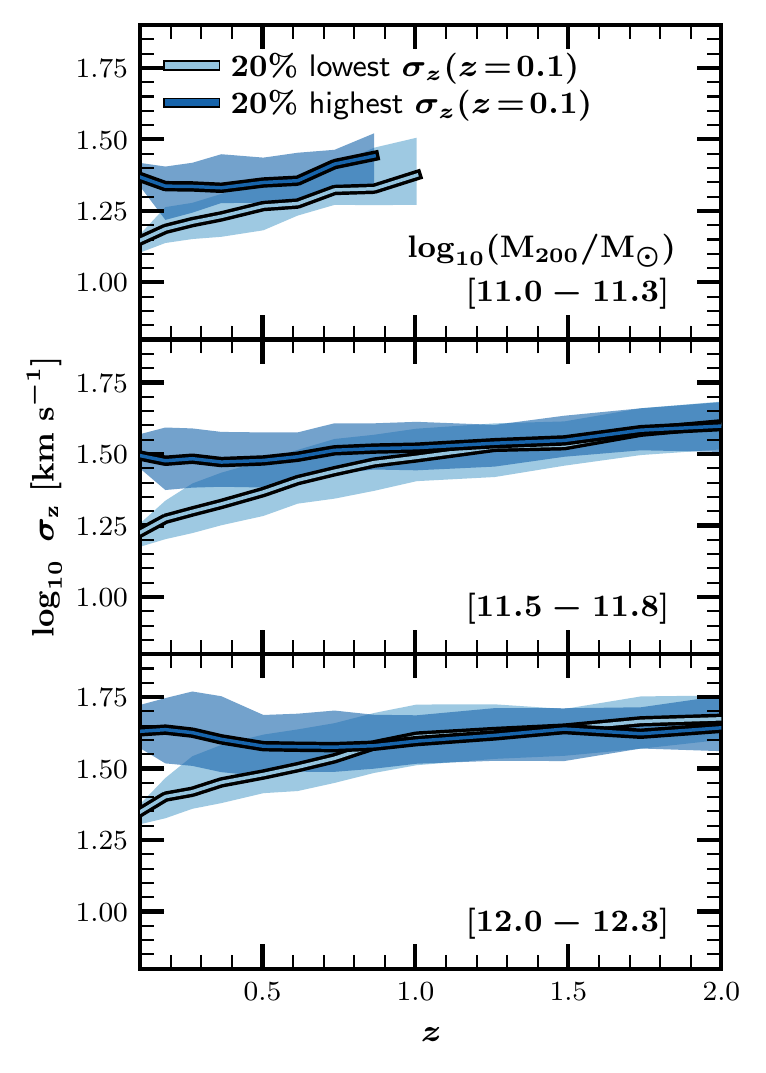}
	\caption{Evolutionary tracks of the gas velocity dispersion for the $z=0.1$ disc galaxies of Fig.~\ref{fig: sigma_profiles}. Different panels correspond to separate bins of $\rm M_{200}$ as indicated by the labels. Dark (light) blue colours refer to galaxies whose $\sigma_z$ at $z=0.1$ is in the top (bottom) quintile of the of the $\sigma_z$ distribution. Solid lines indicate the median evolutionary tracks whereas the shaded regions show the corresponding $16^{\rm th}-84^{\rm th}$ percentiles; both are shown for samples with more than 10 galaxies. }
	\label{fig: sigma_tracks}
\end{figure}

Fig.~\ref{fig: sigma_profiles} shows the $\sigma_z$ profiles computed using equation~(\ref{eq: sigmaz}) for galaxies at $z=0.1$ in {\sc Ref-L100}. The galaxies shown here meet the selection criteria introduced in Section~\ref{sec: GalSamples}; the different panels correspond to different $\rm M_{200}$ bins (see the vertical grey bands in Fig.~\ref{fig: stellar-to-halo-mass rel}). The thick grey lines show the median $\sigma_z(R)$ for all galaxies in each halo mass bin; the dark and light blue lines correspond to those in the highest and lowest $20^{\rm th}$ percentiles of $\sigma_z$, respectively. The dashed lines show the corresponding profiles without the thermal component; the departure from the solid lines is minimal, suggesting that in \eagle, the contribution of the thermal component to gas turbulence is negligible. In all cases, the $\sigma_z$ profiles are approximately flat at $R=3\,R_{50}$, which we define as $R_{\rm flat}$. Note that we verified that using $R_{\rm flat}=1-2\times R_{50}$ does not affect our results. However, using smaller $R_{\rm flat}$ values reduces the number of galaxies in our sample because fewer contain at least 500 cold gas particles within that radius, which is the lower limit we impose when estimating $\sigma_z$. An important feature is that galaxies that have a low (high) value $\sigma_z(R_{\rm flat})$ are characterised by $\sigma_z$ profiles that are consistently low (high) compared to the median across all radii, showing that $\sigma_z(R_{\rm flat})$ is a useful way to characterise the $\sigma_z$ profile using a single number. From this point onward, $\sigma_z$ will refer to the value obtained from equation~(\ref{eq: sigmaz}) using $z_{90}$ and $R_{\rm flat}=3\,R_{50}$ as the height and length of the cylinder, respectively. 

Galaxies in {\sc NoFb-L25} typically have smaller $R_{50}$ values than those in {\sc Ref-L100} at the same $\rm M_{\star}$ (by about a factor of 8) and their $\sigma_z$ profiles flatten at much larger multiples of $R_{50}$, closer to $\approx 10\, R_{50}$. Hence, for {\sc NoFb-L25}, we adopt $R_{\rm flat}= 10\, R_{50}$. We note, however, that none of our results are impacted by the choice of $R_{\rm flat}$ provided it encloses most of a galaxy's cold gas mass.

To assess the correspondence between $\sigma_z(R_{\rm flat})$ and observational estimates, we calculate two additional quantities. Firstly, as gas kinematics is typically inferred from line emission produced in star-forming regions, we substitute the mass of each particle in equation (\ref{eq: sigmaz}) by its SFR, enabling us to derive an SFR-weighted global velocity dispersion for each galaxy. Secondly, we obtain a proxy for the velocity dispersion profile used in observations by computing the local gas turbulence in cylindrical shells of width one physical kpc, i.e. we only use the cold gas particles within each shell to estimate $\sigma_z$. We then interpolate the value of the resulting profile at $R=R_{50}$ to obtain a single value for each galaxy. It is important to note that this method is very susceptible to Poisson noise, especially when the number of particles per shell is small, sometimes reaching only a few tens\footnote{This issue is particularly relevant at low redshifts where galaxy gas fractions can be small.}. To mitigate this effect, we impose a minimum requirement of 50 particles in the bin containing $R_{50}$. Albeit this issue, we find strong correlations (characterised by Spearman correlation coefficients of $\gtrsim 0.7$) between the SFR-weighted $\sigma_z$, the local $\sigma_z$, and the $\sigma_z(R_{\rm flat})$ quantity used in this paper (not shown here). This indicates that, on average, all three measurements provide consistent estimations of gas turbulence. 

In Appendix~\ref{app: Convergence} we present convergence tests that compare results obtained from \eagle{} runs with higher mass and force resolution to those obtained from the Reference model. We find that the evolution of $\sigma_z$ and its dependence on halo mass are in good agreement between the runs suggesting our results are not unduly affected by numerical resolution.

\begin{figure*} 
	\centering 
	\includegraphics[width=0.8\textwidth]{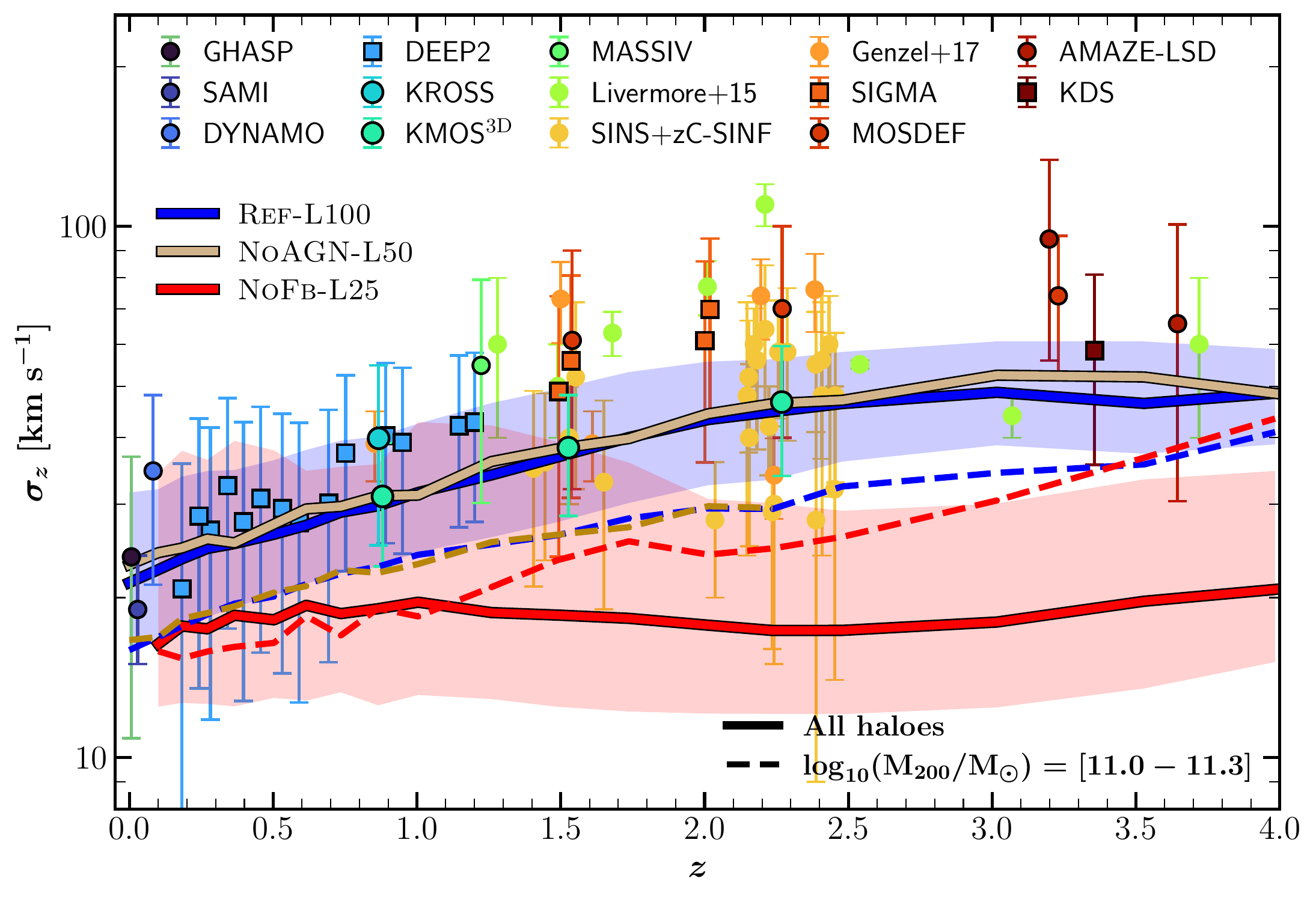} 
	\caption{The evolution of the vertical cold gas velocity dispersion in the {\sc Ref-L100} (blue), {\sc NoFb-L25} (red), and {\sc NoAGN-L50} (light brown) runs, which are compared to the following observational results: {\sf GHASP} \citet{Epinat08, Epinat10}, {\sf SAMI} \citet{Croom12, Varidel20}, {\sf DYNAMO} \citet{Green14}, {\sf DEEP2} \citet{Davis03, Kassin07, Kassin12}, {\sf KROSS} \citet{Stott16, Johnson18}, {\sf KMOS$^{\rm 3D}$} \citet{Wisnioski15, Wisnioski19}, {\sf MASSIV} \citet{Contini12, Epinat12}, \citet{Livermore15}, {\sf SINS+zC-SINF} \citet{FS06, FS09, FS18}, \citet{Genzel17}, {\sf SIGMA} \citet{Simons16, Simons17}, {\sf MOSDEF} \citet{Kriek15, Price20}, {\sf AMAZE-LSD} \citet{Gnerucci11} and {\sf KDS} \citet{Turner17}. Circles (squares) correspond to measurements taken from IFS surveys (long-slit spectroscopy). Symbols with black contours indicate averages over several data points; those without contours correspond to individual measurements. Error bars indicate the mean uncertainties of individual galaxies in each sample, except for {\sf SAMI} where we show the $16^{\rm th}-84^{\rm th}$ percentiles. Note that we only include \eagle~ disc galaxies that are selected as described in Section~\ref{sec: GalSamples}. Solid lines indicate the medians for all \eagle{} galaxies in our samples; shaded regions show the $16^{\rm th}-84^{\rm th}$ percentiles. The dashed lines show \eagle{} results for a fixed halo mass bin, as labelled; note that the $\sigma_{z}$ prediction from all simulations coincide when controlling by halo mass.}
	\label{fig: sigma_ev} 
\end{figure*} 
\begin{figure*} 
	\centering 
	\includegraphics[width=0.85\textwidth]{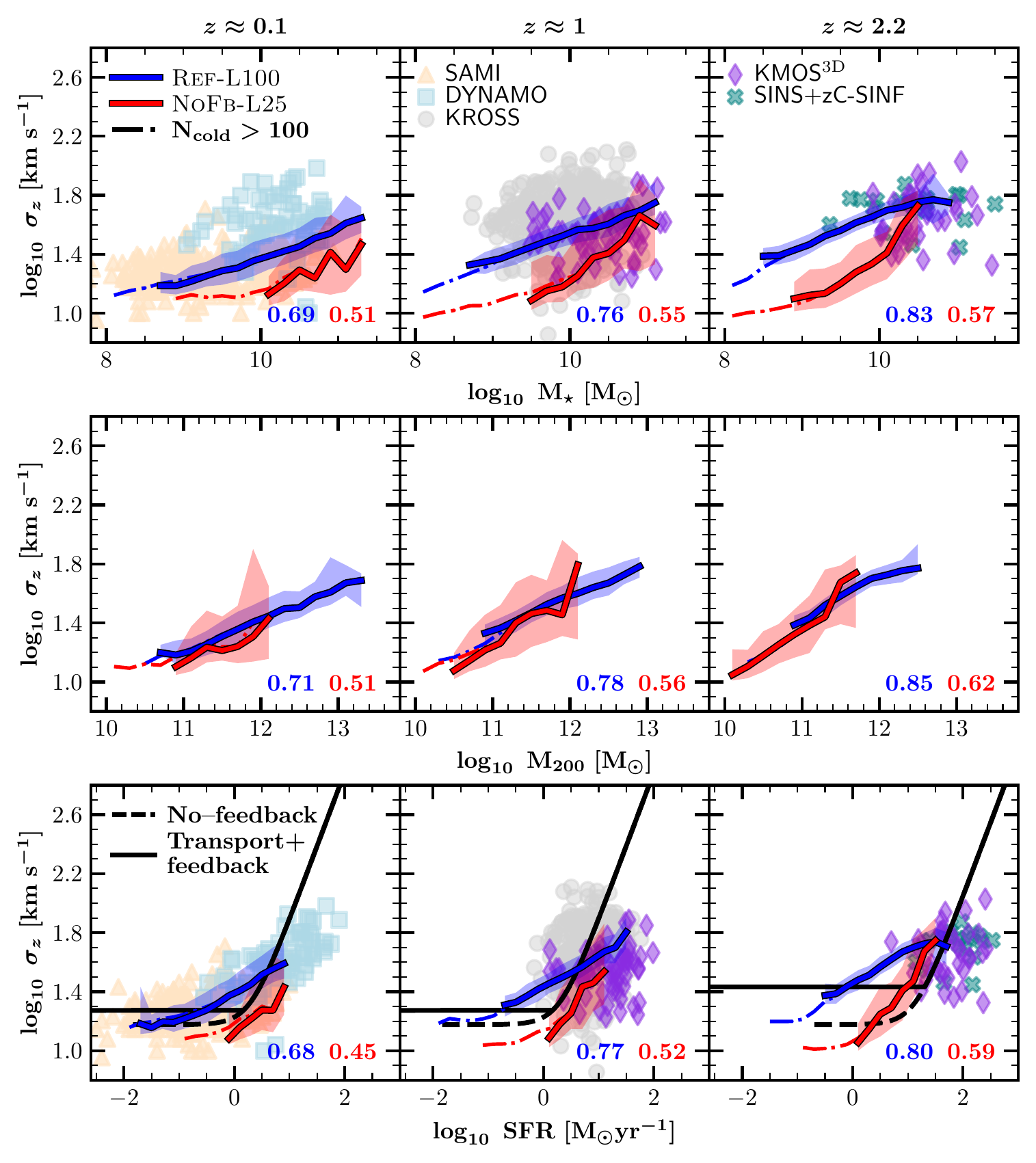} 
	\caption{The vertical velocity dispersion of cold gas plotted as a function of stellar mass (upper), halo virial mass ($\rm M_{200}$; middle) and SFR (bottom) for the {\sc Ref-L100} (blue) and {\sc NoFb-L25} (red) runs. Results are shown for redshifts $\approx 0.1$, 1 and 2.2 (left-hand, middle, and right-hand panels, respectively). Solid lines show the median relations in bins of width $0.3$~dex; shaded regions indicate the $16^{\rm th}$ to $84^{\rm th}$ percentiles. The corresponding Spearman's rank correlation coefficients are shown in the bottom-right corner of each panel (colour coded according to the simulation they pertain to). Dashed-dotted lines show the median trends obtained using a lower minimum threshold of 100 cold gas particles per galaxy to compute $\sigma_{z}$ (solid lines use a minimum of $500$ cold gas particles per galaxy). Individual points are observational results from {\sf SAMI}, {\sf DYNAMO}, {\sf KROSS}, {\sf KMOS$^{\rm 3D}$} and {\sf SINS+zC-SINF} as indicated by the legend. Solid and dashed black lines in the bottom panels show, respectively, the theoretical $\sigma_{z}-{\rm SFR}$ tracks predicted by the transport plus feedback and transport-only models of \citet{Krumholz18}. These theoretical curves were computed using the parameter values {suggested by \citet{Krumholz18} for local spirals (in the bottom-left and bottom-middle panels) and for high-$z$ galaxies (in the bottom-right panel)}.} 
	\label{fig: scaling relations} 
\end{figure*} 

The subsamples of galaxies plotted in Fig.~\ref{fig: sigma_profiles} were chosen to have low and high values of $\sigma_z$ at $R_{\rm flat}$. To investigate the physical origin of these differences, it is important to first verify that they are not short-lived, transient, or stochastic. To investigate this, we analysed the evolution of the vertical gas velocity dispersion for the same subsets of gaseous discs (corresponding to those that lie within the vertical grey bands plotted in Fig.~\ref{fig: stellar-to-halo-mass rel}). As in Fig.~\ref{fig: sigma_profiles}, we select discs whose $\sigma_{z}$ values are in the upper and lower quintile of the $\sigma_z$ distribution (for their halo mass) and track the evolution of $\sigma_z$ in their main progenitors back to $z=2.2$. Note that we only include progenitor galaxies provided their stellar mass exceeds $10^9\,\rm M_{\odot}$, and we eliminate instances in which a galaxy's main progenitor temporarily became a satellite of a more massive halo. Note that we do not, however, impose the $\kappa_{\rm co}>0.7$ selection to the progenitors of the $z=0.1$ galaxies.

Fig.~\ref{fig: sigma_tracks}  shows the median evolutionary tracks of $\sigma_z$ for these galaxies. The median trends (thick solid lines; the shaded regions indicate the 16$^{\rm th}-84^{\rm th}$ percentile scatter) show that the progenitors galaxies with low or high $\sigma_z$ at $z=0.1$ also had low or high $\sigma_z$, respectively, for $z \lesssim 1$. Note that we obtain similar results using the \eagle{} snipshots (which allow us to track $\sigma_z$ every $\approx 100$ Myr). Clearly the differences in $\sigma_z$ at $z=0.1$ (which is due to our selection of galaxies at that redshift) is not a short-lived, transient phenomenon but instead marks a persistent difference between the galaxy samples. The same is true when discs are selected at $z=1$ (not shown here, for clarity). Therefore, any driver of gas turbulence must be acting on long timescales.

% ====================================================================

% =======================================================================
%  SECTION 3: EVOLUTION OF GAS VELOCITY DISPERSION & SCALING RELATIONS
% =======================================================================

\section{Evolution of gas velocity dispersion and scaling relations}
\label{SecEvsigma}

In this section, we present the redshift evolution of the gas velocity dispersion (Section~\ref{subsec: sigma_ev}) as well as some standard scaling relations between $\sigma_z$, $\rm M_{\star}$ and SFR (Section~\ref{subsec: scaling relations}). Throughout the paper we define the SFR as ${\rm SFR}\equiv \Delta m_{\star}/\Delta t$, where $\Delta m_{\star}$ is the total stellar mass formed over a time interval of $\Delta t= 100$ Myr. Using the instantaneous SFR, or the SFR averaged over $\Delta t=10,\, 50$ or $200$ Myr has little impact on our results. 

% =============================================================
\subsection{Evolution of gas velocity dispersion} \label{subsec: sigma_ev}

Fig.~\ref{fig: sigma_ev} shows the evolution of $\sigma_z$ for \eagle~ disc galaxies from $z=0$ to $z=4$ obtained from the {\sc Ref-L100} (blue lines), {\sc NoFb-L25} (red lines) and {\sc NoAGN-L50} runs (light brown lines). A variety of observational data are also shown, including results from long-slit spectroscopy (squares) and IFS surveys (circles; table 5 in \citealt{Ubler19} contains the full compilation of velocity dispersion measurements plotted here). Most references report gas kinematics inferred from H$\alpha$ emission lines, although high-redshift surveys such as {\sf AMAZE-LSD} \citep{Gnerucci11} and {\sf KDS} \citep{Turner17} use [OIII] as the primary target line. Data points with black edges correspond to medians for datasets where multiple measurements are available; error bars indicate the $1\sigma$ scatter. The remaining points from \citet{Livermore15}, \citet{Genzel17} and {\sf SINS+zC-SINF} \citep{FS06, FS09, FS18} correspond to measurements for individual galaxies. Each observational survey aimed to define a sample of rotating discs in the star-forming sequence for which reliable kinematic measurements can be extracted, and various selection criteria were imposed to do so. For example, objects that display either merger or AGN activity are excluded as that can potentially perturb the inferred kinematics. The largest galaxy samples from IFU surveys plotted in this figure correspond to {\sf SAMI} \citep{Varidel20}, {\sf KROSS} \citep{Stott16, Johnson18} and {\sf KMOS$^{\rm 3D}$} \citep{Wisnioski15, Wisnioski19}, which have a homogeneous coverage of the star formation main sequence and cover several orders of magnitude in stellar mass. 

Overall, results from the {\sc Ref-L100} run are in good qualitative agreement with the observational results; the median $\sigma_{z}$ systematically increases from $\approx \rm 20 km\ s^{-1}$ at $z=0$ to $\approx \rm 50\ km\ s^{-1}$ at $z=2.5$, as observed. Even though the fiducial simulation (i.e. the {\sc Ref-L100} run) reproduces the observed evolution of gas turbulence, the similarities with observations should be interpreted with caution. Firstly, to compare with \eagle, we assume that the line-of-sight velocity dispersion ($\sigma_{\rm LoS}$), which is more readily obtained from observations, is comparable to the vertical velocity dispersions we measure from our simulations. In general, however, $\sigma_{\rm LoS}>\sigma_z$ as the line-of-sight component can capture radial and azimuthal motions when galaxies are observed at different inclination angles. Additionally, as mentioned in the previous section, the global $\sigma_z$ used in this paper provides a reasonable estimate of gas turbulence, but it represents a distinct quantity from the local turbulence sometimes inferred from observations. Also, a detailed comparison would also require accounting for other systematics, such as the influence of the simulation's box size, which affects the halo mass range probed, as well as the impact of beam-smearing, which can have an important impact on the observation of high-redshift galaxies. 

We find that the simulation results are similar when the thermal component of $\sigma_z$ (see equation~\ref{eq: sigmaz}) is neglected, indicating that it has a minimal contribution to $\sigma_{z}$ in \eagle{} galaxies on the mass scales we study (see \citealt{Pillepich19} for a different conclusion from the TNG simulation). Note that the gas turbulence inferred by {\sf DYNAMO} (blue circle at $z\approx 0.1$) is considerably higher than that obtained from \eagle. This is due to the survey's selection criteria, which targets analogues of high-redshift galaxies at $z=0.1$, resulting in a sample with systematically higher SFR and gas fractions than galaxies on the main sequence, which are likely to be massive systems with high velocity dispersion.

Note that results obtained from the {\sc NoAGN-L50} run are remarkably similarly to those from {\sc Ref-L100}, suggesting that AGN feedback is not an important driver of gas turbulence on the mass scales probed by our runs (differences may however be evident for more massive systems, of which there are few in the 50 cMpc {\sc NoAGN-L50} run).

Galaxies in the {\sc NoFb-L25} run, however, exhibit a lower $\sigma_{z}$ across all redshifts, with very little evolution. At first glance, this could be taken as an indication that stellar feedback is an important driver of gas turbulence and that it is largely responsible for the systematically higher values of $\sigma_{z}$ and its evolution in the {\sc Ref-L100} run. However, when galaxies are selected according to halo mass we find that all three \eagle{} runs predict a similar evolution of $\sigma_z$. The dashed lines in Fig.~\ref{fig: sigma_ev}, for example, show the evolution of $\sigma_z$ for galaxies occupying haloes with virial masses in the range $10^{11}-10^{11.3}\,{\rm M_\odot}$. The difference between the solid blue and red lines is therefore due to the different halo masses typically occupied by galaxies in each simulation, and in particular the lack of massive haloes in the {\sc NoFb-L25} run (see Fig.~\ref{fig: stellar-to-halo-mass rel}).

%=======================================================================
\subsection{Scaling relations of the gas velocity dispersion} \label{subsec: scaling relations}

The scatter in the $\sigma_{z}$-redshift relation is driven by different galaxy and halo properties. Fig.~\ref{fig: scaling relations} shows the $\sigma_{z}-{\rm M_\star}$, $\sigma_{z}-{\rm M_{200}}$ and $\sigma_{z}-{\rm SFR}$ relations (top, middle and bottom rows, respectively) for both the {\sc Ref-L100} (blue) and {\sc NoFb-L25} (red) runs. The gas turbulence correlates with all three properties, in agreement with correlations reported by observational studies (see observations shown in  Fig.~\ref{fig: scaling relations}). Indeed, the Spearman correlation coefficients ($\rho_{\rm S}$; labelled at the bottom-right corners in each panel) indicate that the strength of the correlation between $\sigma_{z}$ and $\rm M_\star$, and $\sigma_z$ and SFR are significant for both the {\sc Ref-L100} and {\sc NoFb-L25} runs. Note that the lack of high-mass galaxies in the {\sc NoFb-L25} run is due to the smaller simulation box (25 cMpc), whereas the excess of them at low masses is a result of the overproduction of stars due to the absence of feedback.  

At $z \approx 0.1$, galaxies in {\sc Ref-L100} display a $\sigma_{z}-{\rm M_\star}$ relation that resembles the one obtained by the {\sf SAMI} galaxy survey \citep[][shown as tan triangles]{Varidel20}. However, at fixed $\rm M_\star$, the \eagle~ results are systematically lower than the $\sigma_{z}$ measurements obtained for DYNAMO galaxies; this is likely due to the the survey selection effect discussed above. In fact, in the $\sigma_{z}-{\rm SFR}$ plane (bottom-left panel), the {\sc Ref-L100} galaxies show remarkable agreement with both the {\sf SAMI} and {\sf DYNAMO} relations, although the latter are shifted to slightly higher SFRs. Similar agreement is found at higher redshifts. %The {\sf KMOS}$^{\rm 3D}$ galaxies appear to be higher in $\sigma_{z}$ in the top-middle panel compared to the {\sc Ref-L100} run and the {\sf KROSS} observations, but that is simply because KMOS3D galaxies have higher SFRs. 

It is useful to compare results obtained from the {\sc Ref-L100} and {\sc NoFb-L25} runs. The top and bottom rows of Fig.~\ref{fig: scaling relations} suggest that, at fixed $\rm M_\star$ or SFR, the galaxies in the {\sc NoFb-L25} run have systematically lower $\sigma_{z}$ values than those in {\sc Ref-L100}. It is tempting to relate these differences to the absence of feedback in {\sc NoFb-L25}, which is a plausible source of turbulent energy. However, the middle rows of Fig.~\ref{fig: scaling relations} show that this is unlikely the case. Here we plot the $\sigma_{z}-{\rm M_{200}}$ relation for the two runs. In this case, both models follow the same relation. Systematic differences in the cold gas velocity dispersion of galaxies in these runs (at fixed $\rm M_\star$ or SFR) are therefore due to galaxies of comparable stellar mass or SFR occupying haloes of different virial mass. This foretells one of our main results: the thermal energy injected into the ISM by stellar feedback contributes little to turbulent motions in cold gaseous discs. This conclusion is indeed supported by the strong correlation between $\sigma_z$ and SFR in the {\sc NoFb-L25}. Why would galaxies with higher SFRs exhibit higher levels of turbulent motions even in the absence of stellar and AGN feedback? We return to this discussion below. 

% ==============================================================
%             SECTION 4: PHYSICAL DRIVERS OF GAS TURBULENCE
% =============================================================

\section{The physical drivers of gas turbulence} \label{sec: PhysicalDrivers}

In this section we break down the role played by several physical processes in establishing the velocity dispersion in gaseous discs. In Section~\ref{SecGalMergers} we consider the effects of galaxy mergers, in Section~\ref{SecGravInst} we focus on gravitational instabilities, in Section~\ref{SecFeedback} we consider in more detail the impact of stellar feedback, and in Section~\ref{SecGasAcc} the effect of gas accretion.

\begin{figure}
	\centering
	\includegraphics[width=\columnwidth]{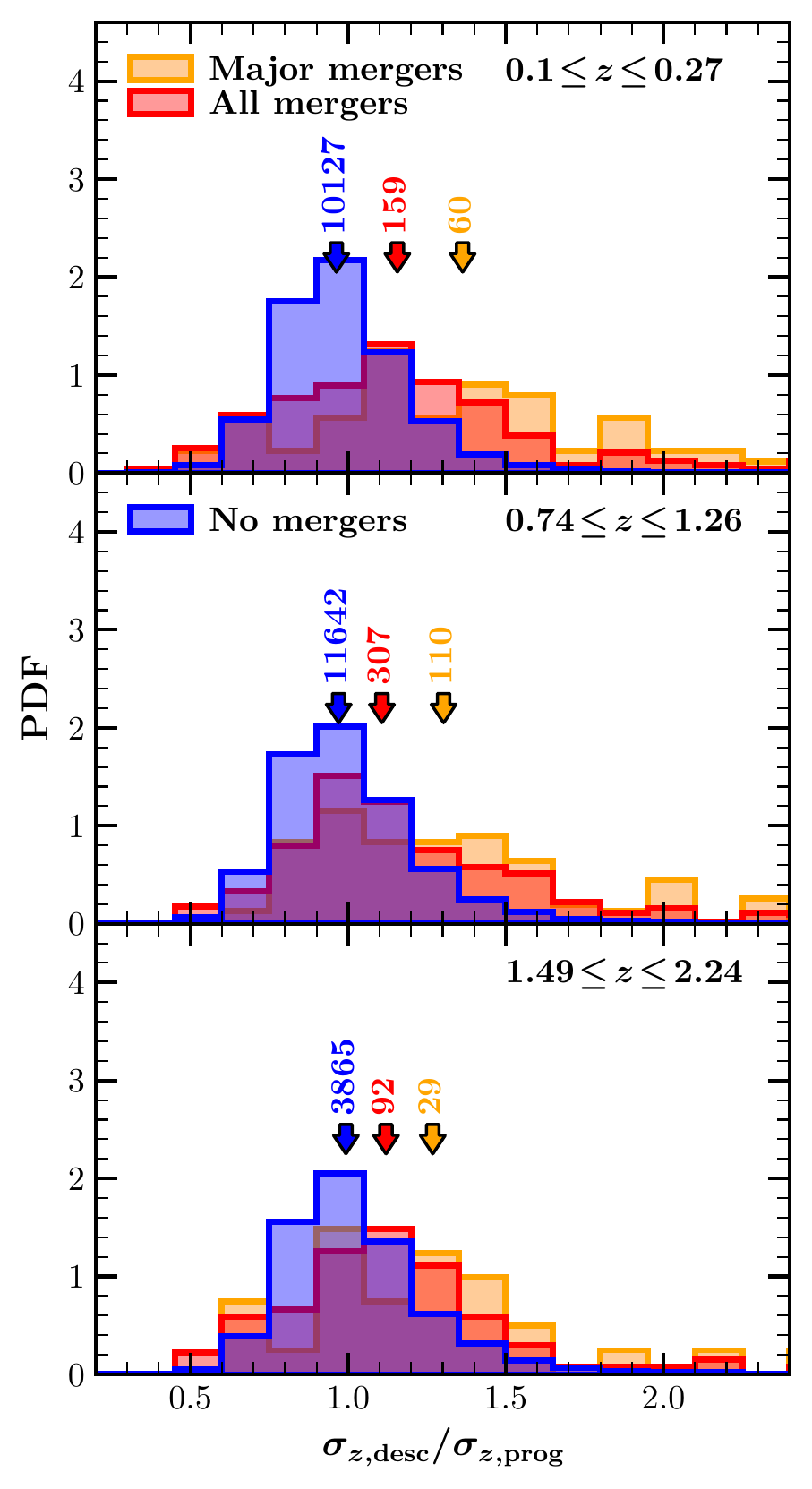}
	\caption{The distribution of the ratio of the gas velocity dispersion of descendants and their main progenitors for galaxies that have not had mergers (blue), for those that have had major mergers (orange), and either major or minor merger (red). Results are shown for the redshift ranges $0.1\leq z\leq 0.27$ (top panel), $0.74\leq z\leq 1.26$ (middle panel) and $1.49\leq z\leq 2.24$ (bottom panel). Downward arrows of corresponding colour indicate the medians of each distribution; the numbers above show the number of events in each sample. Values of $\sigma_{z,{\rm desc}}/\sigma_{z,{\rm prog}}>1$ show that that galaxy {\it increased} its velocity dispersion. On average, mergers increase the vertical velocity dispersion, $\sigma_{z}$.}
	\label{fig: mergers}
\end{figure}
% =======================================================================
\subsection{Turbulence due to galaxy mergers}\label{SecGalMergers}

We link galaxy descendants and progenitors in adjacent snapshots using the galaxy merger trees available in the \eagle\ database \citep{Qu17}. For galaxies that have $\ge 2$ prominent progenitors, we 
follow \citet{Lagos18} and compute the stellar mass ratio between the first and second most massive progenitors to classify merger events. Major mergers are those whose stellar mass ratio is above $0.3$, minor mergers are those with mass ratios between $0.1$ and $0.3$. Mass ratios below $0.1$ are considered ``unresolved'' mergers and, consequently, classed as accretion events \citep[e.g.][]{Crain17}. 
The reason for the latter is that, at the resolution of the \eagle~ simulation, and for galaxies with $\rm M_\star \ge 10^9\,\rm M_{\odot}$, we can only reliably identify mergers with mass ratios $\ge 0.1$. Galaxies at any redshift that went through a major or minor merger in the previous snapshot are added to a ``merger sample'' at that redshift. Conversely, those that have a single progenitor, or multiple progenitors with mass ratios $<0.1$ are included in a ``smooth accretion sample''. 

Fig.~\ref{fig: mergers} shows the Probability Distribution Function (PDF) of the ratios of the gas velocity dispersion in descendants and their most massive progenitors, $\sigma_{z, {\rm desc}}/\sigma_{z, {\rm prog}}$. We show the distributions for galaxies that underwent major mergers (orange), major or minor mergers (red), and galaxies in the smooth accretion sample (blue). Note that we combine results from multiple adjacent snapshots in order to increase the sample sizes. For example, the top panel includes results from redshift pairs $z=0.1-0.18$ and $z=0.18-0.27$. The number of events in each sample are shown on top of the arrows which also indicate the median of the $\sigma_{z, {\rm desc}}/\sigma_{z, {\rm prog}}$ distributions. Note that imposing a $\kappa_{\rm co}>0.7$ could lead to the removal of galaxies whose morphology changes from disc to spheroid as a result of merger (which, as shown in \citealt{Lagos18}, is likely the case in major mergers). To avoid discarding these descendant-progenitor pairs and to account for a potential causal relationship between mergers and changes in $\sigma_{z}$, we only apply the $\kappa_{\rm co}>0.7$ selection to the main progenitor galaxy, but not to their descendants.

Both major and minor mergers increase the gas turbulence in galaxy discs (i.e. $\sigma_{z, {\rm desc}} > \sigma_{z,{\rm prog}}$ in these cases). The effect is strongest for major mergers (orange histograms), although the number of these events is small compared to galaxies in the smooth accretion sample. Typically, major mergers increase gas velocity dispersion by about a factor of 1.3 to 1.4, depending on redshift. 

Note that the PDFs of galaxies in the smooth accretion sample peak at ratios slightly below one (see blue arrows), indicating that, on average, descendants are kinematically colder than their progenitors. This is consistent with the $\sigma_{z}$ evolution analysed in Section~\ref{SecEvsigma}. The impact of galaxy mergers on the cold gas velocity dispersion of discs, as well as its dependence on the merger mass ratio, is worth investigating more thoroughly. We leave this for future work. 

Although galaxy mergers have an important impact on the gas velocity dispersion, they are quite uncommon on the mass (and redshift) scales we study. Hence mergers have only a minor impact on the evolution of gas turbulence when averaged over a population of galaxies. Indeed, we verified that all of the results we have presented so far are unaffected by the removal of merging galaxies. 

% ==============================================================
\subsection{The relation between vertical velocity dispersion and gravitational instabilities}\label{SecGravInst}

\begin{figure*}
	\centering
	\includegraphics[width=0.9\textwidth]{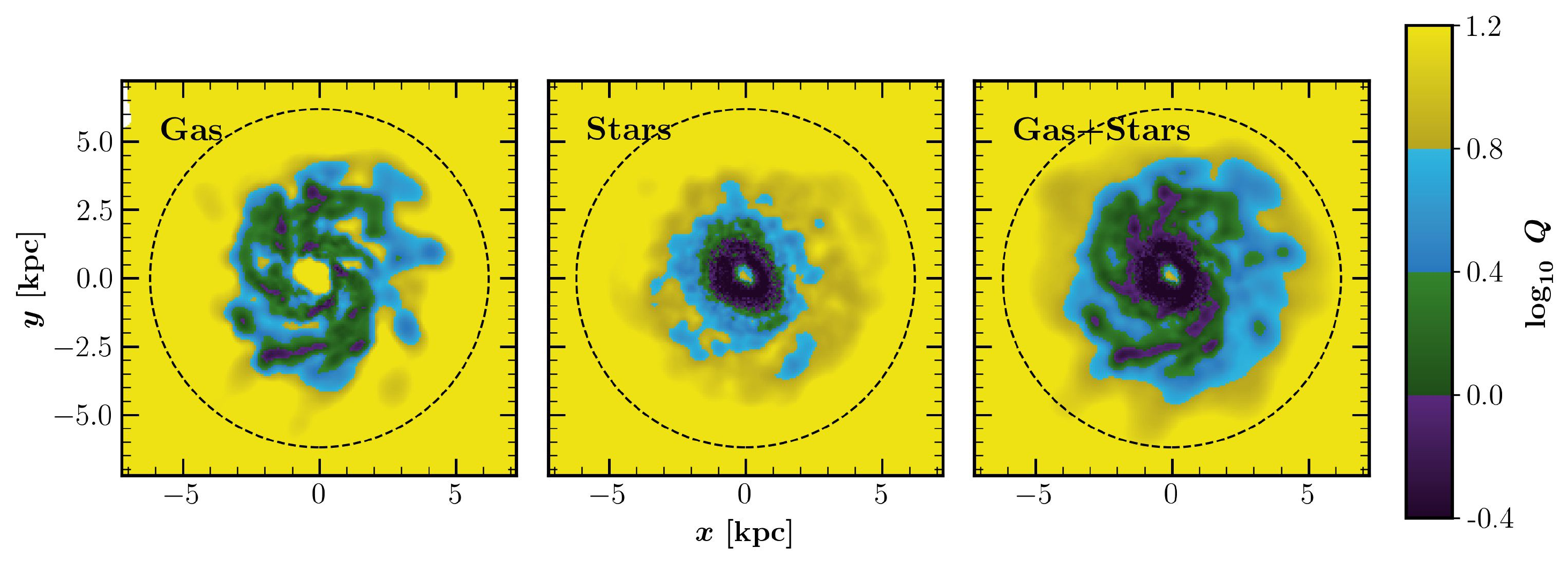}
	\caption{Examples of Toomre-$Q$ instability maps for cold gas (left panel), stars (middle panel), and the combination of cold gas and stars (right panel) for a $z=2$ galaxy identified in the {\sc Ref-L100} run. Colours show different values of $\rm log_{10}(Q)$, as indicated in the colour bar to the right. Green and violet regions are unstable ($Q_{\rm net}<3$). Dashed circles are drawn at $R=3\,R_{50}$, which is the cylindrical aperture within which we measure $\sigma_{z}$.}
	\label{fig: Qmaps}
\end{figure*}

\begin{figure}
	\centering
	\includegraphics[width=\columnwidth]{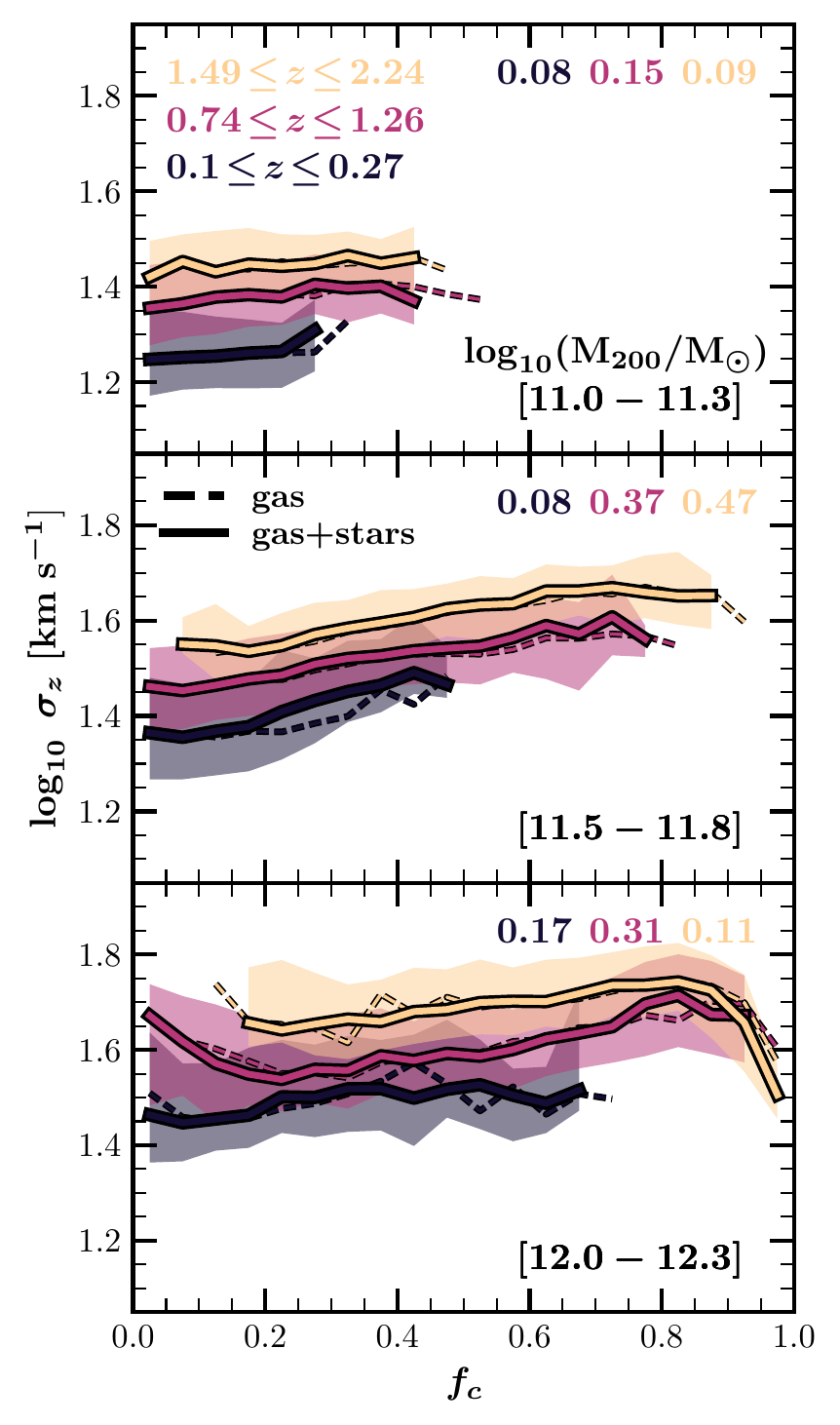}
	\caption{The vertical velocity dispersion of cold gas plotted as a function of the clumpiness parameter, $f_{\rm c}$ (see equation~\ref{eq: clumpiness}), for galaxies in the {\sc Ref-L100} run. Different colours correspond to galaxies identified in different narrow redshift bins, as indicated. Solid lines correspond to the {\em total} baryon mass fraction in clumps, i.e. $f_{\rm c}$ evaluated for both stellar and gaseous particles, whereas dashed lines are evaluated using only the cold gas disc. The shaded regions (only shown along with the solid lines) correspond to the $16^{\rm th}-84^{\rm th}$ percentile ranges. The Spearman correlation coefficients (corresponding to the solid lines) are shown in the top right corner of each panel.}	\label{fig: Clumpiness}
\end{figure}

Observations of discs at high redshift reveal that a significant fraction of their cold gas is contained in distinct self-gravitating clumps \citep[e.g.][]{Elmegreen09, Genzel11}. These clumps may introduce perturbations to the gravitational potential, which can influence the kinematics of baryons in the disc. In particular, non-axisymmetric torques induced by clumps can result in angular momentum loss, creating inward flows of mass while releasing gravitational potential energy. Different analytical models for star-forming discs incorporate this transport mechanism and consider it an important driver of gas turbulence \citep[e.g.][]{Aumer10, Krumholz16}. 

The formation and evolution of self-bound clumps can be studied using the theory of gravitational instabilities. In this framework, local instabilities arise due to an imbalance between gravity and restoring forces and are commonly quantified using the Toomre stability parameter \citep{Toomre64}. This can be defined for gaseous or stellar discs as
\begin{equation} \label{eq: Toomre-Q}
 Q_i = \frac{\kappa\, \sigma_{r,i}}{\pi\, G \,\Sigma_i}.
\end{equation}
where $i$ refers to the disc's baryonic component\footnote{Note that for the stellar component $\pi$ is replaced by 3.36}, $\kappa$ is the epicyclic frequency, $\sigma_{r,i}$ is the radial velocity dispersion, and $\Sigma_i$ the surface density. Note that the quantities in equation~(\ref{eq: Toomre-Q}) are typically measured locally; the value of $Q_i$ can therefore vary from place to place within the disc. Typically, regions where $Q_i>Q_{\rm crit}$ are considered stable against collapse, with $Q_{\rm crit}\approx 1$ indicating marginal stability. 

Some analyses account for the destabilising effects of stars and the stabilising effect of the finite disc thickness \citep[e.g.][]{Romeo13}. For our analysis, we account for instabilities in both the stellar and gaseous disc. Following \citet{Inoue16}, we compute 2D maps of $Q_{\rm gas}$ and $Q_\star$ using face-on projections of our \eagle~ galaxy sample, and use the formulation of \citet{Romeo11} to calculate the multi-component $Q$ stability parameter, which we denote $Q_{\rm net}$. We refer to Appendix~\ref{ComputingQ} for a description of how we construct $Q_{\rm net}$ maps from maps of $Q_{\rm gas}$ and $Q_\star$. In Appendix~\ref{AppTestingQ}, we report tests of our method's accuracy using cylindrically-symmetric galaxy models for which the $Q_\star(r)$ profiles are known. 

Fig.~\ref{fig: Qmaps} shows the $Q_i$ maps obtained for a $z=2$ disc galaxy identified in the {\sc Ref-L100} run. To disentangle the contribution from gas and stars, we show $Q_{\rm gas}$ and $Q_\star$ maps separately (left and middle panels, respectively), although for our analysis we only use the $Q_{\rm net}$ maps (right panel; which contain contributions from both gas and stellar particles). In general, the stellar component is the primary driver of instabilities in the galaxy's inner regions, whereas gas dominates the instabilities in the outer disc, as shown in this example. The same is true for the majority of discs used in our analysis. In general, the $Q_{\rm net}$ maps contain more unstable regions than either of the individual $Q_i$ maps, suggesting that neither the stellar nor the gaseous component should be excluded, especially in cases where they make similar contribution to local instabilities. This is important as often $Q_{\rm gas}$ alone is used to interpret the observations of gas velocity dispersion. However, we note that the contribution from stars becomes increasingly important at low redshifts, when the typical gas fractions of discs are lower. 

To assess whether gravitational instabilities drive gas turbulence, we first define a ``clumpiness'' parameter, $f_{\rm c}$, as
\begin{equation} \label{eq: clumpiness}
    f_{\rm c} = \frac{{\rm M_{bar}}(Q_{\rm net}<3)}{\rm M_{\rm bar}},
\end{equation}
where ${\rm M_{\rm bar}}(Q_{\rm net}<3)$ is the baryonic mass (of stars plus gas) contained within pixels with $Q_{\rm net}<3$, and $\rm M_{\rm bar}$ is the total baryonic mass of the galaxy. Both quantities are evaluated using the particles enclosed within the same cylindrical aperture that was used to compute $\sigma_z$ (see Section~\ref{sec: sigma_z calculation}), which typically encloses the majority of unstable regions in the discs. The threshold $Q_{\rm net}<3$ used to select unstable regions is motivated by the work of \citet{Inoue16}, who studied high redshift galaxies in ``zoom-in'' cosmological simulation and found that instabilities are prone to form at locations where $Q_{\rm net} \lesssim 2-3$ \citep[see also][]{Elmegreen11}.  

In principle, galaxies with high $f_{\rm c}$ values are those that may also contain a large number of self-bound clumps which causes the mass flows through the disc and subsequently increase the gas turbulence. Since $f_{\rm c}$ is designed to trace these unstable regions, it is reasonable that the parameter is also associated, to some extent, with an increase in turbulence. Fig.~\ref{fig: Clumpiness} shows $\sigma_{z}$ as a function of the clumpiness parameter for galaxies in different bins of redshift (the same bins used for Fig.~\ref{fig: mergers}) and virial mass (the values of $\rm M_{200}$ increase from the top to bottom panels). Solid lines show $f_{\rm c}$ values calculated using all baryons in the disc (as defined in equation~\ref{eq: clumpiness}) whereas dashed lines use only the cold gas component (i.e. $f_{\rm c}={\rm M_{gas}}(Q_{\rm net} < 3)/{\rm M_{gas}}$). Our definition of clumpiness is less accurate when most of the baryons within the aperture are in regions with $Q_{\rm net}<3$. When this occurs in \eagle, galaxies appear as one large unstable ``clump" rather than as a disc containing multiple small self-bound clumps. This explains the non-monotonic behaviour at $f_{\rm c} \gtrsim 0.8$, where $\sigma_z$ mainly traces the kinematics of baryons within unstable regions, which are intrinsically cold. In addition to our analysis, we also investigate the relationship between the classical clumping factor, $C = \left<\rho^2\right>/\left<\rho\right>^2$ and gas turbulence (here $\rho$ correspond to the mass density of the gas). Our findings (not shown here) reveal that these two parameters are uncorrelated across all halo masses and redshifts examined in this study. Therefore, we conclude that $f_{\rm c}$ serves as a more appropriate indicator of gravitational instabilities.

\begin{figure*}
	\centering
	\includegraphics[width=0.9\textwidth]{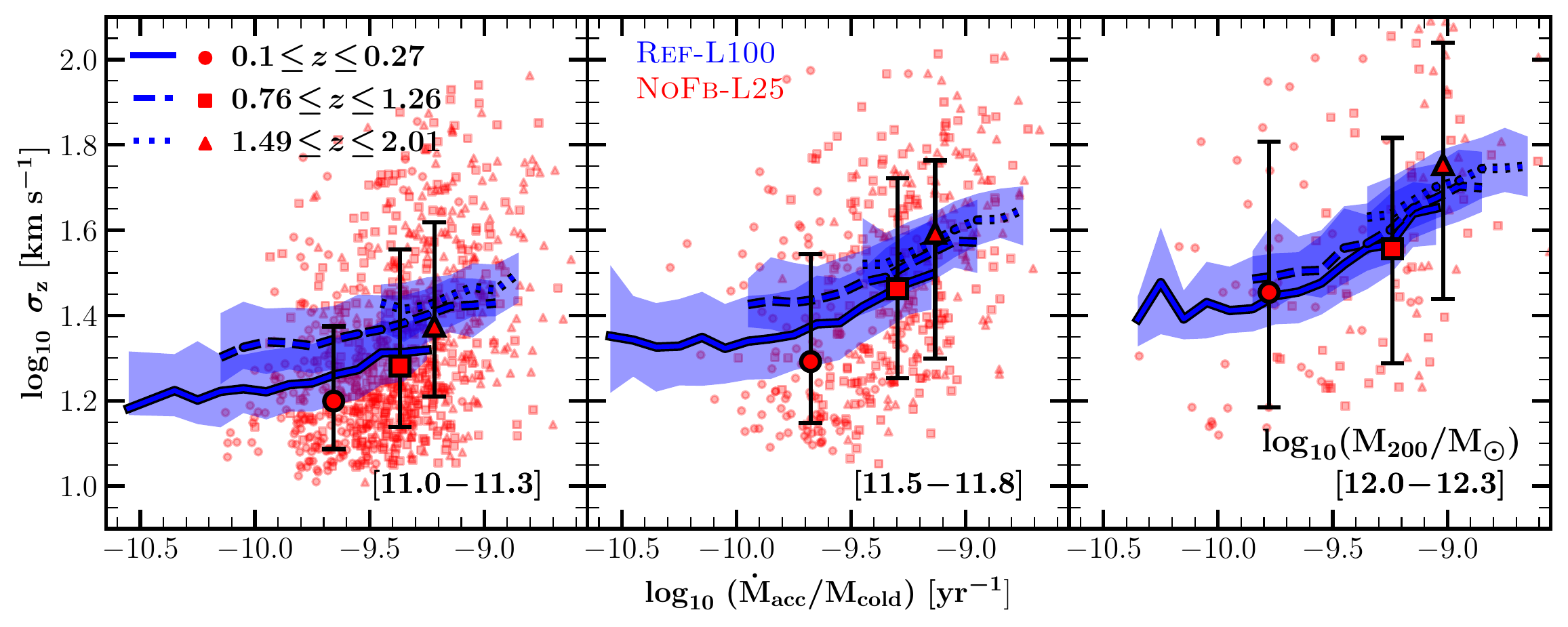}
	\caption{Vertical gas velocity dispersion plotted as a function of the specific gas accretion rate for galaxies hosted by low (left panel), intermediate (middle panel) and high mass haloes (right panel), as indicated in brackets in the bottom-right corners of each panel, in the {\sc Ref-L100} (blue) and {\sc NoFb-L25} (red) runs. Different line styles and symbols correspond to results from different redshift ranges, as indicated by the legend. Lines represent median relations, and shaded regions indicate the $16^{\rm th}$ and $84^{\rm th}$ percentiles in {\sc Ref-L100}. Red symbols in the background show individual galaxies from the {\sc NoFb-L25} run, whereas outsized symbols and error bars indicate the median and $16^{\rm th}$ and $84^{\rm th}$ percentiles for galaxies identified at different redshifts, respectively. }
	\label{fig: accretion}\end{figure*}

Overall, the correlations between $\sigma_{z}$ and $f_{\rm c}$ are quite weak at most redshifts and halo masses we analysed. Consider, for example, the lowest halo mass bin (top panel), for which the Spearman rank correlation coefficients (the coloured numbers labelled in the upper-right corner) are less than $0.12$ at all redshifts. For intermediate masses, and at redshifts $z\gtrsim 0.74$, the correlation between $\sigma_{z}$ and $f_{\rm c}$ is somewhat stronger, but it weakens again for higher halo masses. The fact that, at fixed $\rm M_{200}$, the $\sigma_{z}$ depends more strongly on redshift than on $f_c$ suggests that the evolution of gas turbulence is only weakly related to $f_{\rm c}$, at least for galaxies in the \eagle~ simulation.

% ==============================================================
\subsection{Stellar Feedback-driven turbulence}
\label{SecFeedback}

The observed correlation between $\sigma_{z}$ and SFR could imply that stellar feedback is an important driver of gas turbulence: a higher SFR implies a higher abundance of massive stars whose supernovae inject energy into the ISM, leading perhaps to a higher $\sigma_z$. As shown in Fig.~\ref{fig: scaling relations}, \eagle\ predicts $\sigma_z-{\rm M_\star}$ and $\sigma_z-{\rm SFR}$ relations that are in reasonable agreement with the observed relations at several different redshifts. In particular, there is a relatively weak relation between $\sigma_{z}$ and SFR among systems with low SFRs (which is particularly clear at low redshifts), but this transitions to a stronger dependence for galaxies with high SFRs.

\citet[][hereafter K18]{Krumholz18} developed an analytical model for the evolution of gaseous discs assuming hydrostatic equilibrium and marginal gravitational instability.  In their model, supernovae feedback and the radial transport of gas driven by gravitational instabilities both inject turbulent energy to the ISM, while energy is dissipated by shocks on a timescale comparable to the local crossing time. These energy sources are assumed to reach equilibrium; this model is referred to as ``Transport$+$Feedback'' (see bottom panels in Fig.~\ref{fig: scaling relations}). We also show the K18 ``No-feedback'' model, which only accounts for the effect of radial transport within the discs, but not for the turbulent energy injected by supernovae.  In order to compare their predictions with our \eagle\ results, we adopt the parameter values suggested by K18 to best describe spiral galaxies at $z\le 1$ (bottom left and middle panels in Fig.~\ref{fig: scaling relations}) and  high$-z$ galaxies at $z=2$ (bottom right panel in Fig.~\ref{fig: scaling relations}). Because the K18 model only accounts for the neutral (atomic plus molecular) gas component, we add $15\ {\rm km\ s^{-1}}$ (in quadrature) to their results \citep[see][]{Krumholz16} when comparing with our \eagle\ predictions (which include all ISM components). 

Even though, qualitatively, there are common features between the {\sc Ref-L100} run and the predictions of K18's fiducial model, it is clear that in detail the two differ. For example, the flattening of $\sigma_z$ at low SFRs is less abrupt in \eagle; the \eagle\ trend is perhaps better described as a weak dependence of $\sigma_{z}$ on SFR at $\rm SFR<1\,\rm M_{\odot}\,yr^{-1}$. Only after relaxing the condition of having a minimum of 500-cold gas particle to compute $\sigma_z$ (see Section~\ref{sec: GalSamples}) does a flattening becomes apparent (see dotted-dashed lines, for which we instead impose a minimum number of cold gas particles per galaxy of 100), albeit at much lower SFRs ($\lesssim 0.1\,\rm M_{\odot}\,yr^{-1}$) than predicted by K18. At higher SFRs, \eagle\ predicts a weaker dependence of $\sigma_{z}$ on SFR than K18's fiducial model. When comparing the {\sc NoFb-L25} run with the K18 ``No-Feedback'' model we also see clear differences, with \eagle\ predicting lower $\sigma_z$ than the minimum values in the K18 model (which is higher than the floor value of $15\ {\rm km\ s^{-1}}$ described above). 
%The differences between \eagle\ and K18 could be taken as an indication that stellar feedback is at best only partially driving $\sigma_z$ in \eagle.

Considering again the comparison between {\sc Ref-L100} and {\sc Nofb-L25}, the most significant result of the $\sigma_{z}-{\rm SFR}$ correlation is that in both runs galaxies with higher SFRs also have a higher $\sigma_z$. In {\sc NoFb-L25}, however, there is no injection of energy into the ISM by supernovae, suggesting that the higher gas velocity dispersions associated with high SFRs in that run are {\em not} related to feedback. Feedback therefore cannot be the only driver of turbulence, and perhaps not even the most important one.

The $\sigma_{z}-{\rm M_{200}}$ relations for {\sc Ref-L100} and {\sc NoFb-L25}, shown in the middle panels of Fig.~\ref{fig: scaling relations}, are remarkably similar at overlapping masses. This implies that the differences seen in the other relations (specifically $\sigma_z-{\rm M_\star}$ and $\sigma_z-{\rm SFR}$) arise because galaxies of the same stellar mass or SFR occupy haloes of very different virial mass. In this context, for a fixed $\rm M_\star$ or SFR, gas particles with a particular kinetic energy are more likely to become unbound and escape the halo in the {\sc NoFb-L25} run, as they live in lower mass haloes than galaxies of the same stellar mass in {\sc Ref-L100}. This could result in systematically lower values of $\sigma_{z}$ in {\sc NoFb-L25} at fixed ${\rm M_\star}$ or SFR, potentially leading to the offset between the two runs seen in the upper and lower panels of Fig.~\ref{fig: scaling relations}. 

It is important to highlight that, even though the $\sigma_z-{\rm M_{200}}$ relations are similar in the two runs, at fixed halo mass the scatter in $\sigma_{z}$ is larger in the {\sc NoFb-L25} run than in the {\sc Ref-L100} run (which is reflected in the lower values of the Spearman rank correlation coefficients in {\sc NoFb-L25}). A possible interpretation of this result is that stellar feedback acts as a regulation mechanism that leads to a tighter relation between $\sigma_{z}$ and ${\rm M_{200}}$. When feedback is absent so is the regulation mechanism, resulting in a broader distribution in $\sigma_{z}$ values at fixed halo mass. This is reminiscent of the effect of stellar feedback on the SFR$-\rm M_\star$ plane, and the relationship between SFR, stellar mass, and molecular gas explored in \citet{Lagos16}. Those authors showed that by making stellar feedback stronger (weaker), the relation between these quantities became tighter (broader).

The similarity of the $\sigma_{z}-{\rm M_{200}}$ relations in the {\sc Ref-L100} and {\sc NoFb-L25} runs indicates that stellar feedback is not the primary driver of gas turbulence in \eagle. The positive correlation between $\sigma_{z}$ and SFR may therefore be due to a more fundamental driver of turbulence; one which increases the SFRs of galaxies along with their velocity dispersion. In addition to the impact of either mergers or dynamical instabilities (quantified by the $f_{\rm c}$ ``clumpiness'' parameter) -- which, as shown above, cannot fully explain the evolution of $\sigma_z$ -- we consider next the impact of cosmological gas accretion and how it affects the dynamics of cold gas discs.

% =====================================================================
\subsection{Gas Accretion-driven turbulence}\label{SecGasAcc}

%Recent studies  have shown that accretion onto galaxy discs can contribute to gas turbulence \citep[e.g][]{Ginzburg22, Forbes23}. Similar to inward gas flows in Toomre-unstable discs, gas from the CGM accretes onto the galaxy by releasing gravitational potential energy. The latter is transferred to the disc and can potentially imprint features in the kinematics and morphology. In particular, if the accreted gas mass is high, it can enhance perturbations in the gas distribution, such as an increase in the disc scale height.

\begin{figure}
	\centering
	\includegraphics[width=\columnwidth]{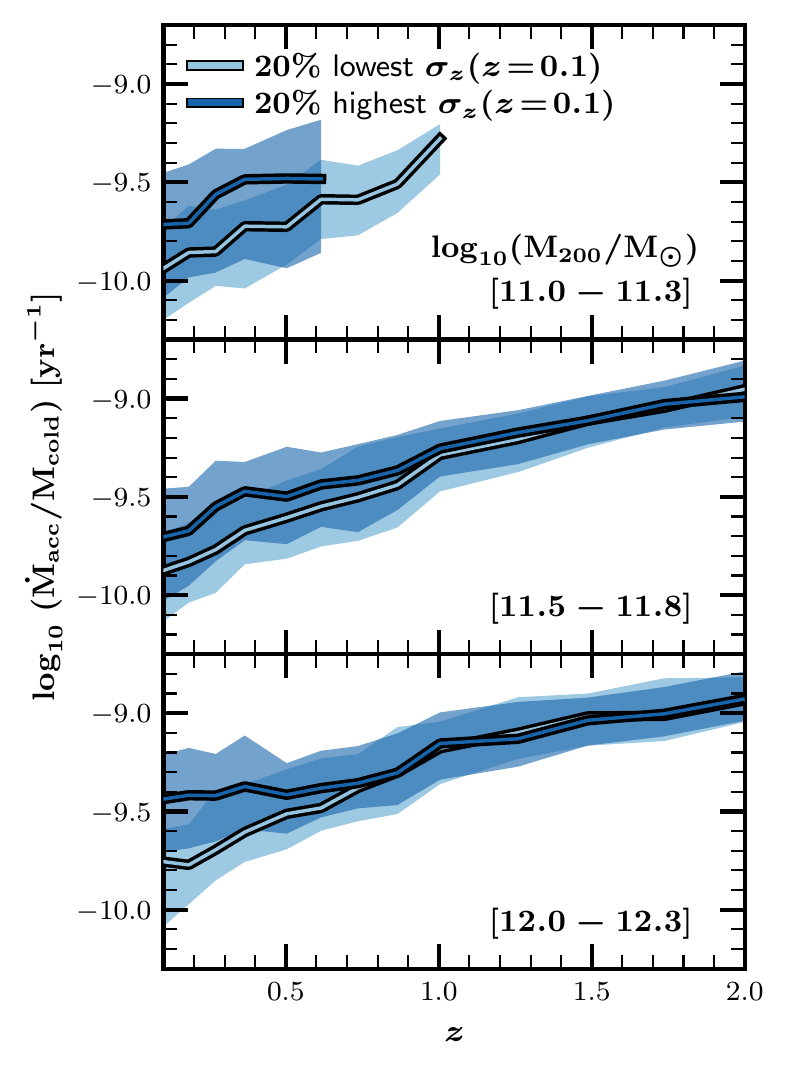}
	\caption{Same as Fig.~\ref{fig: sigma_tracks}, but showing the evolutionary tracks of the specific accretion rate, $\dot{\rm M}_{\rm acc}/ \rm M_{cold}$, for disc galaxies identified at $z=0.1$. Dark lines and shaded regions correspond to galaxies whose $\sigma_z$ values (at $z=0.1$) are in the upper quintile for their halo mass; lighter colours correspond to galaxies in the lower quintile. Note that the evolution of $\dot{\rm M}_{\rm acc}/\rm M_{cold}$ -- specifically the separation of the different evolutionary tracks -- resembles that of $\sigma_{z}$ shown in Fig.~\ref{fig: sigma_tracks}, suggesting that the two are closely connected.}
	\label{fig: sMacc_tracks}
\end{figure}

In Fig.~\ref{fig: accretion} we plot $\sigma_{z}$ as a function of the specific gas accretion rate, i.e. $\dot{\rm M}_{\rm acc}/{\rm M_{cold}}$, for galaxies that lie in a few separate bins of ${\rm M_{200}}$ and for three different redshift bins. Results are shown for both the {\sc Ref-L100} and {\sc NoFb-L25} runs (blue and red colours, respectively). Clearly $\sigma_{z}$ correlates with the specific gas accretion rate, although the trend varies with halo mass and redshift. For all halo masses, the $\sigma_{z}-\dot{\rm M}_{\rm acc}/\rm M_{cold}$ relation flattens when accretion rates are low ($\dot{\rm M}_{\rm acc}/\rm M_{cold}\lesssim 10^{-9.8}\ yr^{-1}$), while at higher specific accretion rates the relation steepens. At fixed halo mass, higher specific accretion rates also occur at higher redshifts.%, reflecting the fact that these halos are rarer and originate from higher linear rms density fluctuations.

\begin{figure*}
	\centering
	\includegraphics[width=0.85\textwidth]{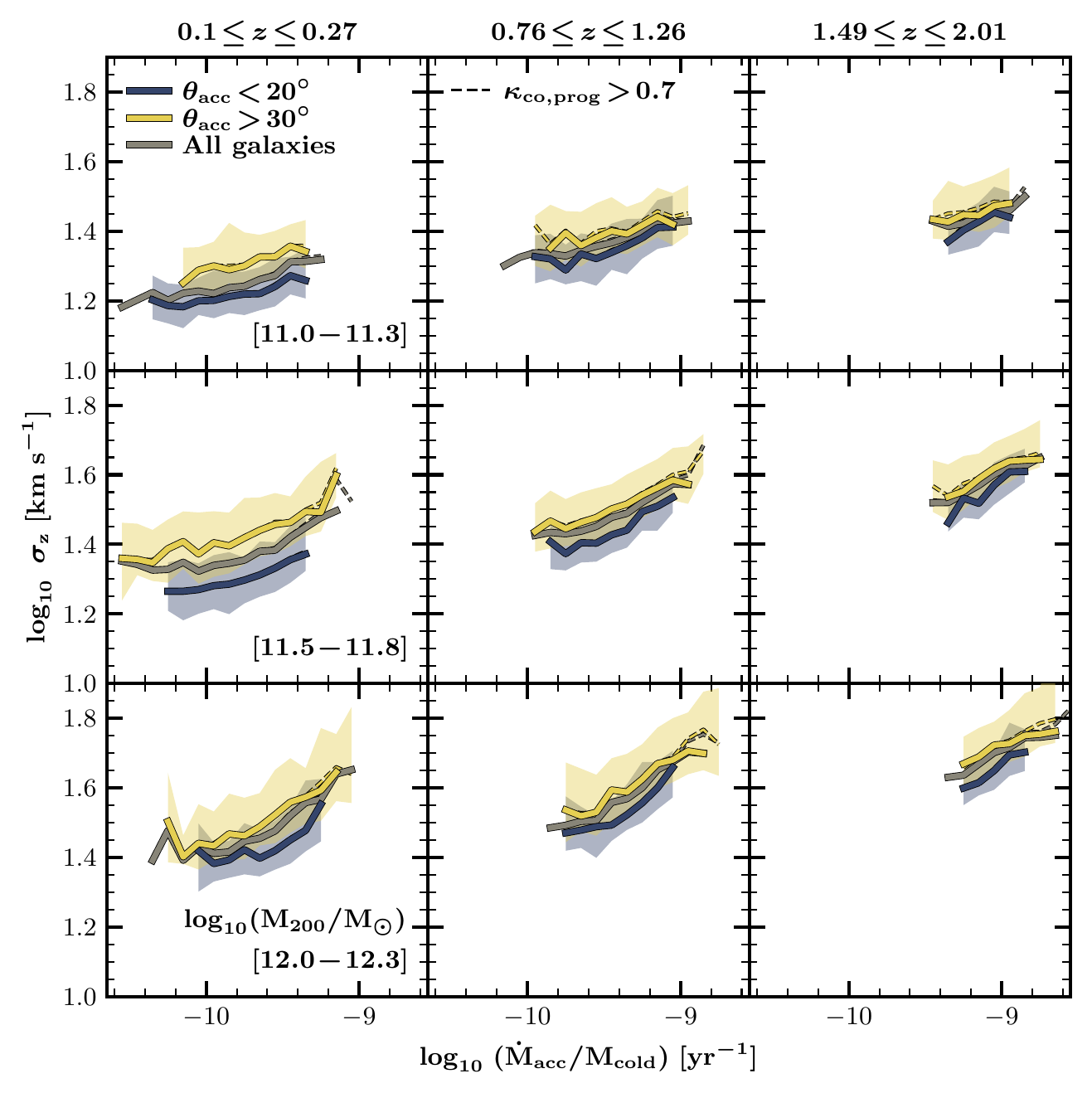}
	\caption{The vertical gas velocity dispersion plotted as a function of the specific accretion rate for galaxies in the {\sc Ref-L100} run identified at low (left), intermediate (middle), and high (right) redshifts, and hosted by low (top), intermediate (middle) and high (bottom) mass haloes, as labelled. Results are shown for the full sample (grey lines) and for subsamples for which the angular momentum of the accreting gas is closely-aligned (blue lines) or misaligned (yellow lines) with the rotation axis of the gaseous disc. Solid lines refers to samples containing discs only (i.e. $\kappa_{\rm co}>0.7$ for cold gas particles) whereas dashed lines refers to samples where only progenitors are required to meet the $\kappa_{\rm co}>0.7$ criterion.}
	\label{fig: misalignment}
\end{figure*}

%The vertical gas velocity dispersion (normalised by $V_{200}$) plotted as a function of the specific accretion rate for galaxies in the {\sc Ref-L100} run identified at low (left panel), intermediate (middle panel), and high (right panel) redshifts. Results are shown for the full sample (grey lines) and for subsamples for which the angular momentum of the accreting gas is closely-aligned (blue lines) or misaligned (yellow lines) with the rotation axis of the gaseous disc. Solid lines refers to samples containing discs only (i.e. $\kappa_{\rm co}>0.7$ for cold gas particles) whereas dashed lines refers to samples where only progenitors are required to meet the $\kappa_{\rm co}>0.7$ criteria.

At fixed halo mass and redshift, there are fewer galaxies in the {\sc NoFb-L25} run than in {\sc Ref-L100}, so for the former we use outsized points with errors bars to indicate the median values of $\sigma_z$ and $\dot{\rm M}_{\rm acc}/\rm M_{cold}$ (error bars indicate the 16$^{\rm th}$ to 84$^{\rm th}$ percentile scatter in $\sigma_z$). By comparing these points to the blue lines in each panel, however, it is clear that results obtained from the {\sc NoFb-L25} run are similar to those obtained from {\sc Ref-L100}. Indeed, at fixed halo mass and redshift, both simulations predict similar values of $\sigma_{z}$ for the same specific accretion rates, indicating that this relation is likely more fundamental than the relationship between $\sigma_z$ and SFR. The specific accretion rate of cold gas may therefore be the primary driver of turbulence. Note, however, that the scatter in $\sigma_z$ is considerably larger in the {\sc NoFb-L25} run than it is in {\sc Ref-L100}, which is perhaps not surprising given the large scatter in the $\sigma_z-{\rm M_{200}}$ relation seen in that run (see the middle panels of Fig.~\ref{fig: scaling relations}). The increased variability of $\sigma_z$ in the {\sc NoFb-L25} may be interpreted as an {\it indirect} consequence of turning feedback off, which eliminates the various regulatory mechanisms mentioned in Section~\ref{SecFeedback} and aligns with results from \citet{Wright20}, who showed that feedback is effective in regulating the amount of matter that can be accreted onto haloes (and consequently, onto galaxies). The evolution of disc kinematics will then be a result of the interplay between outflows (generated by stellar feedback) and inflows. In the absence of feedback, these regulatory mechanisms do not exist, which we speculate leads to the larger scatter in the $\sigma_z-{\rm M_{200}}$ and $\sigma_z -\dot{\rm M}_{\rm acc}/\rm M_{cold}$ relations. 

The connection between $\sigma_z$ and accretion rate is also evident in the redshift evolution of $\dot{\rm M}_{\rm acc}/\rm M_{cold}$ and $\sigma_z$ of individual galaxies. We use the \eagle\ mergers trees to track the evolution of the specific gas accretion rates onto discs selected at $z=0.1$ for the same bins of virial mass used to construct Fig.~\ref{fig: sigma_tracks}. The results are shown in Fig.~\ref{fig: sMacc_tracks}. Note that the evolution of $\dot{\rm M}_{\rm acc}/\rm M_{cold}$ resembles that of $\sigma_{z}$: galaxies with low (high) velocity dispersion at $z=0.1$ exhibited lower (higher) specific accretion rates for an extended period of time prior to this. Comparing Fig.~\ref{fig: accretion} with Fig.~\ref{fig: sigma_tracks}, it is clear that the evolutionary tracks of the specific accretion rate for the two galaxy samples converge at around the same cosmic time as their $\sigma_{z}$ values do (at $z\approx 1$ and $\approx 0.7$ for the intermediate and high $\rm M_{200}$ bins, respectively).

As mentioned above, however, $\sigma_z$ is largely independent of accretion rate when the latter is low, but there remains considerable scatter. This motivates the study of the effects of {\em misaligned} gas accretion, which we quantify using the angle $\theta_{\rm acc}$ between the net angular momentum vector of the disc and that of the accreting cold gas. As for Fig.~\ref{fig: accretion}, we focus on galaxies identified at low, intermediate, and high redshifts. At each redshift, we split our entire sample of galaxies into those with mostly aligned ($\theta_{\rm acc}<20^{\circ}$) and mostly misaligned ($\theta_{\rm acc}>30^{\circ}$) gas accretion. %Note that simultaneously binning by $\theta_{\rm acc}$, halo mass, and redshift reduces the size of our galaxy sample significantly. We therefore do not bin by halo mass, but instead normalise the vertical velocity dispersion by $V_{200}$, thus removing the $\rm \sigma_z-{\rm M_{200}}$ relation.  
Fig.~\ref{fig: misalignment} shows the $\sigma_z-\dot{\rm M}_{\rm acc}/\rm M_{cold}$ relation for the full galaxy sample (solid grey lines), as well as for the aligned (solid blue lines) and misaligned (solid yellow lines) subsamples. Different panels correspond to different redshift ranges (increasing from left to right, as in Fig.~\ref{fig: accretion}) and to different halo mass bins (increasing from top to bottom). Dashed lines show the equivalent relations when the descendant galaxies with $\kappa_{\rm co}<0.7$ are also included (note that their progenitors are still discs with $\kappa_{\rm co}>0.7$; i.e. in this case we include galaxies whose accretion history may have driven them to non-disky morphologies -- note that this selection only affects galaxies that are rapidly accreting). 

It is clear from Fig.~\ref{fig: misalignment} that the scatter in the $\sigma_z-\dot{\rm M}_{\rm acc}/\rm M_{cold}$ relation is closely connected to the alignment of the disc and accreting material: when accreting gas is misaligned, discs tend to be more turbulent. Moreover, there are hints that the impact of misaligned accretion on $\sigma_{z}$ is more important for galaxies with low specific accretion rates, which are more common at low redshifts. Note that when including galaxies with $\kappa_{\rm co}<0.7$ (dashed lines), the relation between $\sigma_z$ and $\dot{\rm M}_{\rm acc}/\rm M_{cold}$ becomes more evident at high specific accretion rates. This suggests that, in this regime, accretion can significantly perturb the disc's structure, driving many to non-disky morphologies.

In summary, Figs.~\ref{fig: sigma_tracks},~\ref{fig: accretion}~and~\ref{fig: misalignment} show that the rate and geometry of gas accretion both play an important role in establishing the vertical velocity dispersion, $\sigma_z$, of \eagle\ galaxies. 

% ==============================================================
%       SECTION 5: DISCUSSION 
% =============================================================

\section{Discussion} \label{sec: Discussion}

\begin{figure}
	\centering
	\includegraphics[width=\columnwidth]{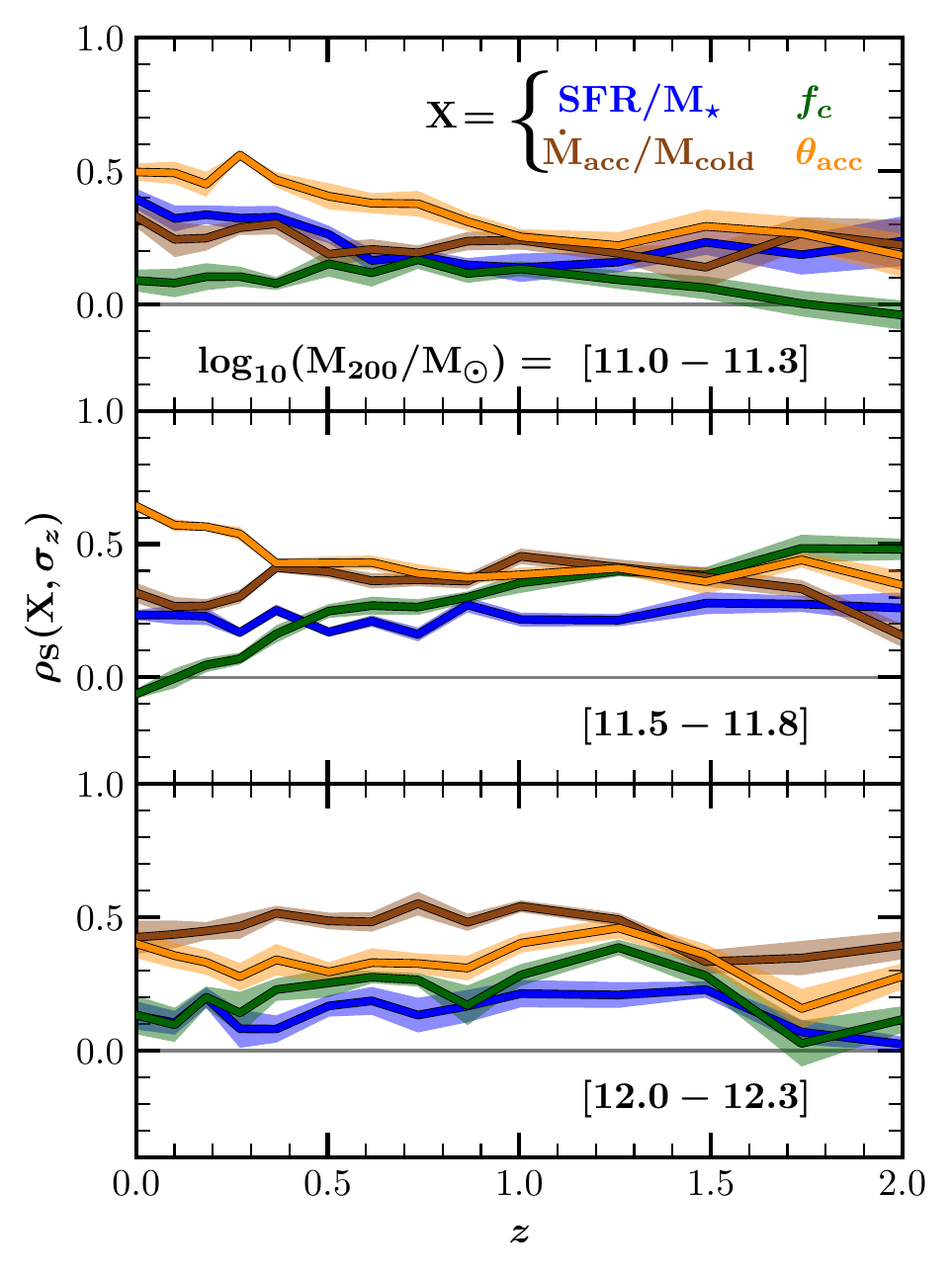}
	\caption{Spearman rank correlation coefficient, $\rho_{\rm S}$, for the $\sigma_{z}-{\rm sSFR}$ (blue), $\sigma_{z}-f_{\rm c}$ (green),  $\sigma_{z}-{\dot{\rm M}_{\rm acc}/ \rm M_{cold}}$ (brown) and $\sigma_{z}-\theta_{\rm acc}$ (orange) relations as a function of redshift and for galaxies occupying low (top), intermediate (middle) and high (bottom) mass haloes. At all masses and redshifts, all samples contain at least 100 galaxies. The shaded regions represent the errors estimated from 10 jackknife subsamples (see text for details). Note that the thick lines show the $\rho_{\rm S}$ values obtained from the entire population of galaxies within each sample; they are not obtained from the merger trees of galaxies identified at a particular time. The correlation between $\sigma_{z}$ and the different galaxy properties depends on both redshift and the mass of the host haloes. At most redshifts and halo masses, the geometry and rate of accreting cold gas correlates most strongly with $\sigma_z$.}
	\label{fig: Spearman}
\end{figure}

\subsection{How important is each physical driver across redshift and halo mass?}
 
The correlations between $\sigma_{z}$, (specific) SFR, the clumpiness factor, $f_{\rm c}$, and (specific) gas accretion rates, in addition to the correlations between the three latter properties, supports the widely adopted assumption that galaxy discs evolve as a consequence of quasi-stable equilibrium between inflows, star formation and outflows (e.g. \citealt{Forbes14a, Rathaus16, Wang22}). In this context, an increase in the inflow rate will deposit a larger amount of gas into the disc, which in turn will induce local instabilities across the disc which collapse into dense clouds and increase the SFR. Depending on the mass of the galaxy, the subsequent feedback-driven outflows will expel gas and suppress the SFR, lowering the self-gravity of the gaseous disc. The latter helps to stabilise the disc against fragmentation ($Q\gtrsim 1$).

Fig.~\ref{fig: sigma_tracks} suggests that turbulent kinetic energy is not injected stochastically but rather occurs on prolonged timescales, allowing a galaxy to maintain a high or low $\sigma_{z}$ for several Gyr. Using the \eagle{} snipshots, we verify that $\sigma_z$ does not exhibit short timescale ($\approx 100$~Myr) trends with either SFR or specific gas accretion rate, which supports the argument of a constant injection of turbulence over long periods of time.  There is an interesting similarity between the ``memory'' of $\sigma_{z}$ for \eagle\ galaxies and the memory of their position in the ${\rm SFR}-\rm M_{\star}$ plane. For example, \citet{Matthee19} selected galaxies along the star formation main sequence and showed that galaxies that lie above the mean relation typically stay above for several Gyr. The same is true for galaxies below the main sequence. They also find a halo-mass dependence of the timescale in which galaxies retain ``memory'' of their position relative to the main sequence. Similarly, \citet{Lagos17} and \citet{WaloMartin20} showed that, in \eagle, the progenitors of $z=0$ galaxies with low and high stellar spin display differences in their stellar kinematics for several Gyr, only to become indistinguishable at $z\gtrsim 1$, an effect that has also been found in other simulations (e.g. \citealt{Penoyre17, Choi17}). These results are related to how haloes assemble, and in the case of \citet{Lagos17} and \citet{WaloMartin20} the main distinction between the $z=0$ high and low stellar spin galaxies was found to be star formation in the former and sustained gas accretion in the latter. In the context of the drivers of $\sigma_z$, Fig.~\ref{fig: sMacc_tracks} suggests that the evolution of accretion rates offers an explanation for the long-timescale memory of the gas turbulence.  

Depending on the physical conditions of discs (and those of their host haloes), the processes affecting the velocity dispersion in gaseous discs can have different degrees of importance at different redshifts or halo masses. To quantify the relationship between $\sigma_{z}$ and the various galaxy properties explored in this paper, we computed the Spearman correlation coefficient, $\rho_{\rm S}$, for each of these relations. This approach does not disentangle which physical process causes high-velocity dispersion but provides quantitative insight into the main drivers of the scatter in the $\sigma_{z}-{\rm M_{200}}$ relations (see middle panels in Fig.~\ref{fig: scaling relations}) at different redshifts, and complements the detailed analysis of previous sections. Fig.~\ref{fig: Spearman} shows the evolution of $\rho_{\rm S}$ for the $\sigma_{z} - {\rm sSFR}$ (blue curves), $\sigma_{z}-f_{\rm c}$ (green curves), $\sigma_{z}-\dot{\rm M}_{\rm acc}/{\rm M_{cold}}$ (brown curves) and $\sigma_{z}-\theta_{\rm acc}$ (yellow curves) relations as a function of redshift and for three halo mass bins. Note that we do not track a given population of galaxies through their merger trees, but rather show the values of $\rho_{\rm S}$ obtained for the selected samples at each redshift (see Section~\ref{sec: GalSamples}). We estimate the errors in $\rho_{\rm S}$ via jackknife resampling using galaxies contained within 10 equal but distinct sub-volumes. Thus, each jackknife subsample contains all galaxies in the simulation box but excludes those whose COP is within a specific subvolume. We calculate $\rho_{\rm S}$ for the 10 jackknife subsamples from which we obtain the mean  $\rho_{\rm S}$ and its standard deviation (shown as solid lines and shaded regions, respectively). 

Overall, we find positive correlations (i.e. $\rho_{\rm S}>0$) between $\sigma_{z}$ and the four properties mentioned above. For the $\sigma_{z}-{\rm sSFR}$ relation, $\rho_{\rm S}$ is typically of order $0.1-0.3$ for most redshifts, and reaches a maximum (of roughly 0.4) at low redshifts and low halo masses. The evolution of $\rho_{\rm S}$ for the other relations shows a clear mass-dependence. The median $\rho_{\rm S}$ for the $\sigma_{z}-f_{\rm c}$ relation goes from $\rho_{\rm S} \approx 0.05$ at low masses (see top panel of Fig.~\ref{fig: Clumpiness}) to a weak but positive correlation ($\rho_{\rm S} \approx 0.25$) at intermediate and high halo masses ($\rho_{\rm S}>0.4$ for the intermediate mass bin at high redshifts). Similarly, for the $\sigma_{z}-\dot{\rm M}_{\rm acc}/{\rm M_{cold}}$ the median $\rho_{\rm S}$ goes from $0.23$ to $0.48$, for the low- and high-mass bins, respectively. We find that the correlation with misaligned gas accretion angle is the strongest for low-redshift galaxies living in low- to intermediate-mass haloes. This correlation is weaker at higher redshifts and becomes comparable in strength to that of the $\sigma_{z}-\dot{\rm M}_{\rm acc}/{\rm M_{cold}}$ relation. These conclusions are robust, even after considering the errors in $\rho_{\rm S}$, which are generally small for all relations.

Fig.~\ref{fig: Spearman} shows that the correlation between gas turbulence and specific gas accretion is significant across all redshifts and halo masses. In particular, for Milky-way haloes ($\rm M_{200}\approx 10^{12}\, M_{\odot}$), the typical values of $\rho_{\rm S}$ obtained for the $\sigma_z-\dot{\rm M}_{\rm acc}/{\rm M_{cold}}$ (brown line in bottom panel) is higher than that of the sSFR or $f_{\rm c}$ relations at all times. This suggests that the vertical velocity dispersion of cold gaseous discs within these haloes is governed primarily by the specific gas accretion rate, especially at $z\lesssim 1$ when the $\rho_{\rm S}$ from both sSFR and $f_{\rm c}$ are $\approx 2$ times smaller than that from the specific gas accretion rate. Furthermore, Fig.~\ref{fig: Spearman} indicates that the misalignment of the inflowing gas is also important, and actually plays a important role in setting the gas turbulence for low-$z$ discs within low-mass halos. This is consistent with Fig.~\ref{fig: misalignment} which shows that the segregation of the low and high $\theta_{\rm acc}$ samples is larger at lower redshifts. For higher specific accretion rates, which are more common at high redsfhits (see the differences in the dynamical ranges in the different panels of Fig.~\ref{fig: misalignment}), the segregation of the low- and high-$\theta_{\rm acc}$ samples is smaller, hence the correlation between gas turbulence and $\theta_{\rm acc}$ is expected to be weaker. 

\citet{Sales12} used the {\tt GIMIC} simulation \citep{Crain09} to show that galaxies are very likely to become discs by $z=0$ if at the time of the maximum halo expansion (i.e. the turnaround radius), the spin of their inner halo regions is aligned with the outer halo parts (which contain the gas that will eventually be accreted to the galaxy). 
Our results show that large gas accretion misalignments can lead to more turbulence, which in the more extreme cases affect the morphology of galaxies (see Fig.~\ref{fig: misalignment}). In the future, we will investigate the link between gas turbulence, misaligned gas accretion and galaxy morphology \citep[see also][]{Hafen22}.

Now turning our attention to gravitational instabilities, Fig.~\ref{fig: Spearman} shows that the correlation between $\sigma_z$ and $f_{\rm c}$ become important at $z\gtrsim 1$, and for galaxies in the intermediate halo mass bin. This suggests that, at this epoch, transport-driven turbulence contributes to $\sigma_z$ (see Section~\ref{SecGravInst}).  We note that correlations with the clumpiness parameter are clearer when there is a significant fraction of galaxies with $f_{\rm c}>0.3$. This indicates that gravity-driven turbulence may become efficient when at least a third of the baryons of the discs are in Toomre-unstable regions. Indeed, we find weak or no correlation for discs at low-$z$, which in general have $f_{\rm c}<0.3$ (see dark violet lines in Fig~\ref{fig: Clumpiness}). This is consistent with local discs being mostly Toomre-stable (i.e. $Q_{\rm net}>1$); therefore, inward gas flows along the disc are expected to be small. 

We interpret the positive correlation between $\sigma_{z}$ and ${\rm sSFR}$ mainly as a consequence of the correlation between $\sigma_{z}$ and the specific gas accretion rate, although, depending on redshift and mass, trends with clumpiness or misaligned accretion could also contribute, and even dominate in the latter case. This is plausible: the SFR is connected to the amount of cold gas mass in the disc which in turn depends on the amount of gas accreted, which is, in our interpretation, what is injecting turbulence. Furthermore, Figs.~\ref {fig: scaling relations} and \ref{fig: accretion} suggest that the relationship between gas turbulence and specific gas accretion rate is more fundamental than its relation to SFR because the latter are seen even in the {\sc NoFb-L25} run. 

Previous works based on hydrodynamical simulations have shed light on the possible drivers of gas turbulence, focusing on isolated discs  \citep{Renaud21, Ejdetjarn22}, cosmological zoom-in simulations from {\tt FIRE} \citep{Hung19}, or single galaxies from the TNG50 simulation \citep{Forbes23}. These have shown that in some regimes, it is clear how disc instabilities, or gas accretion, can affect disc turbulence. Our work contributes to this wealth of literature by attempting to identify the drivers of gas turbulence across a broad range of redshifts and galaxy masses. The large samples of galaxies provided by the \eagle\ simulation allowed us to understand that different physical drivers of turbulence may dominate at different times and in haloes of different masses, highlighting the complexity of the problem. The relatively high $\rho_{S}$ values presented in Fig~\ref{fig: Spearman} further suggest that all the processes studied in this paper should be incorporated in explicit models \citep[e.g.][]{Forbes23} that aim to identify the primary causes of high gas turbulence. Furthermore, the evolutionary tracks of $\sigma_{z}$ in \eagle\ reveal that high gas turbulence is a consequence of physical drivers acting on timescales of several Gyr, which disfavours stochastic processes, like the SFR, significantly impacting $\sigma_z$. 

% ===================================================

\subsection{Interpreting the impact of stellar feedback} 

Our selection of cold gaseous discs, as described in Section~\ref{sec: GalSamples}, naturally excludes the gas that has been recently heated by feedback, which for the most part is ejected from the cold regions of the ISM into the CGM in the form of galactic outflows. Hence, $\sigma_z$ mostly traces cold gas that has not been directly affected by the most recent feedback episodes. This resembles most observational studies of gas turbulence, which focus on the connection between the kinematic properties of gas within the disc (excluding gas in outflows) and global galaxy properties such as stellar mass or SFR. Nevertheless, the outflowing gas is expected to interact with the ISM and to deposit into it some of the initial SN energy. Note that, in contrast with several other cosmological simulations, \eagle\ does not turn off the hydrodynamics for SNe-heated gas particles, implying that part of the energy from SNe can be imprinted into the disc and potentially manifest in the form of gas turbulence. This imprint should be encapsulated, to some extent, by the $\sigma_z$-SFR correlations reported here, where the SFR can be thought of as a proxy for the feedback energy that was transferred to the gas in the ISM. 

The \eagle{} prescription for modelling stellar feedback was not designed to deposit turbulence into the ISM, but instead to regulate the SFR by ejecting gas from the ISM. As mentioned in Section~\ref{sec: GalSamples}, this is implemented using a thermal-stochastic scheme that increments the internal energy of gas particles surrounding the SNe. Another technique to generate outflows in cosmological simulations is via a kinetic mode in which feedback energy is injected into fluid elements by increasing their velocities. By construction, these velocity kicks deposit turbulence into the ISM\footnote{Provided the outflowing gas is allowed to interact hydrodynamically with the rest of the cold gas}, which in turn suppresses the formation of dense gas clumps. The latter provides an additional mechanism to regulate star formation. 

\citet{Chaikin23} recently presented a hybrid prescription for SN feedback that implements both the thermal and kinetic channels. Using simulations of isolated Milky-Way galaxies, they showed that the injection of low-energy velocity kicks increase the gas velocity dispersion in the ISM while (as for \eagle{}) the thermal channel primarily ejects cold gas from the disc. Even though introducing a kinetic channel increases gas turbulence, disentangling its contribution to $\sigma_z$ from other physical drivers (such as gas accretion) and investigating its long-timescale effect will require implementing this approach in a cosmological hydrodynamical simulations.

Previous work based on hydrodynamical simulations has shown that clustered SNe can generate more powerful galactic winds than randomly-distributed SNe \citep{Fielding18, Martizzi20}; this presumably strengthens the correlation between SNe feedback and gas turbulence. However, assessing the detailed effects of clustered feedback in \eagle{} is unfeasible due to its limited resolution. Particularly, the temperature floor imposed on gas particles inhibits the formation of dense gas clumps which are the potential sites of clustered feedback. As a consequence, the role of feedback in driving gas turbulence could be underestimated in our analysis. Our results should therefore be interpreted in the context of randomly distributed feedback which is traced by the global SFR of the galaxy.

% ==============================================================
%       SECTION 6: CONCLUSIONS 
% =============================================================

\section{Conclusions} \label{sec: Conclusions}

Using the \eagle~ simulations, we carried out a comprehensive analysis of the vertical velocity dispersion of cold gas, $\sigma_{z}$, in central galaxies in the redshift range $0<z\lesssim 4$. We considered galaxies with rotationally supported gas discs with stellar masses $\rm M_\star \ge 10^{9}\,\rm M_{\odot}$ that also contain at least 500 cold gas particles. The main aims of our paper were to establish whether galaxies in \eagle\ have a similar $\sigma_z$ evolution as that in observed systems and to additionally understand the physical drivers of $\sigma_{z}$. Our analysis focused on the Reference \eagle\ ({\sc Ref-L100}) run, which was supplemented by runs that eliminate stellar and AGN feedback ({\sc NoFb-L25} and {\sc NoAGN-L50}, respectively). Our main results are as follows. 

\begin{itemize}

\item The redshift evolution of $\sigma_{z}$ in {\sc Ref-L100} is in good qualitative agreement with observations from various spectroscopic surveys. Results obtained from the {\sc NoAGN-L50} run, which does not include AGN feedback but does include stellar feedback, were similar to those obtained from the {\sc Ref-L100} run, indicating that AGN feedback is not an important driver of gas turbulence in the mass and redshift range of our analysis (see Fig.~\ref{fig: sigma_ev}). 

\item The relationship between $\sigma_{z}$ and various global galaxy properties are qualitatively similar to those obtained observationally: $\sigma_{z}$ increases with increasing $\rm M_\star$ and SFR, but the dependence weakens below $\rm M_\star \lesssim 10^{9.5}\rm\, M_{\odot}$ (or SFRs $\lesssim 1\,\rm M_{\odot}\, yr^{-1}$). This is consistent with analytic models of the $\sigma_{z}-{\rm SFR}$ relation. However, at high SFRs, \eagle\ predicts a much weaker trend {between $\sigma_{z}$ and SFR than these analytic models}. $\sigma_{z}$ also correlates strongly with the virial mass of a galaxy's host halo, regardless of redshift or of the feedback implementation (see Fig.~\ref{fig: scaling relations}). 

\item We analysed the $\sigma_{z}-{\rm M_\star}$  and $\sigma_{z}-{\rm SFR}$ relations in a run that does not include either AGN or stellar feedback ({\sc NoFb-L25}), and found that, at fixed halo mass, the evolution of $\sigma_z$ mimics that obtained for galaxies in the {\sc Ref-L100} run (Fig.~\ref{fig: sigma_ev}). In addition, the $\sigma_{z}-{\rm M_{200}}$ relations are remarkably alike in these two runs, albeit with a larger scatter in {\sc NoFb-L25}. The $\sigma_z-{\rm M_{\star}}$ and $\sigma_{z}-\rm SFR$ relations in {\sc NoFb-L25} have a similar slope but lower normalisation than they do in {\sc Ref-L100}.% However, the fact that, even in the absence of feedback, a positive correlation between $\sigma_{z}$ and $\rm SFR$ is still present (see Fig.~\ref{fig: scaling relations}, suggests that in \eagle\ stellar feedback is not the primary driver of gas turbulence in discs.

\item The evolutionary tracks of $\sigma_{z}$ for individual
galaxies resemble the evolution of their specific gas accretion rates (see Fig.~\ref{fig: sigma_tracks} and \ref{fig: sMacc_tracks}, respectively). The evolutionary tracks of $z=0.1$ discs of low and high $\sigma_{z}$ continue to be systematically different up to $z \approx 1$. The exact timescale over which this ``memory'' of a high or low $\sigma_{z}$ value is preserved depends on halo mass and the cosmic time at which we select galaxies, but it persists provided there are systematic differences in their gas accretion rates.
%If we look at discs selected at $z=1$ and trace their progenitors, a very similar $\sigma_{z}$ ``memory'' is obtained, indicating that $\sigma_{z}$ does not evolve stochastically on short timescales. 

\item On average, galaxy mergers increase the gas velocity dispersion in discs, with major mergers causing the most significant increase. Over the mass and redshift range we consider, however, the number of galaxies affected by merger is small and has little impact on the global evolution of $\sigma_{z}$. 

\item At some mass scales, we find a weak correlation between $\sigma_z$ and the fraction of mass contained in regions susceptible to collapse by gravitational instabilities (referred to as the clumpiness parameter, $f_{\rm c}$). In our approximate treatment, $f_{\rm c}$ aims to quantify how transport mechanisms that originate from clump formation affect gas turbulence. The $\sigma_{z}-f_{\rm c}$ correlations are significant at redshifts $z\gtrsim 1$, as well as in halos with virial mass $\rm M_{\rm 200} \gtrsim 10^{11.5}\, M_{\odot}$, consistent with gravitational instabilities being important in discs with high gas fractions (typical at high redshifts). 

\item At fixed halo mass, $\sigma_{z}$ correlates with the specific gas accretion rate, $\dot{\rm M}_{\rm acc}/\rm M_{cold}$. The trends, however, weaken when the accretion rate is low, but become strong for galaxies with $\dot{\rm M}_{\rm acc}/{\rm M_{cold}} \gtrsim \rm 10^{-9.5}\ yr^{-1}$, which are common at high redshifts and among massive haloes. Similar results are obtained from the {\sc NoFb-L25} run, although in this case the $\sigma_{z}-\dot{\rm M}_{\rm acc}/\rm M_{cold}$ relations exhibit larger scatter. The tighter $\sigma_{z}-\dot{\rm M}_{\rm acc}/\rm M_{cold}$ correlation in the {\sc Ref-L100} run may be a manifestation of regulatory feedback mechanisms, which are not present in the {\sc NoFb-L25} (hence the large scatter). 

\item Misaligned gas accretion, quantified by the angle between the disc's rotation axis and the angular momentum vector of accreting cold gas, correlates strongly with $\sigma_z$ in haloes with low specific accretion rates, i.e.  $\dot{\rm M}_{\rm acc}/{\rm M_{cold}} \rm \lesssim 10^{-9.5}\ yr^{-1}$, which typically correspond to haloes with virial masses $\rm M_{\rm 200}<10^{11.8} \, M_{\odot}$ at $z\lesssim 1$. Thus -- in the regime of low gas accretion rate, where $\sigma_{z}$ is largely independent of the {\em net} rate of accretion -- turbulence is higher among galaxies whose accreted gas is misaligned with respect to the disc's angular momentum.
\end{itemize}

%Our results suggest that a complex interplay of different physical processes determine the level of turbulence in cold gas disc's. The relative importance of each mechanism depends on the halo mass and the epoch at which a galaxy is identified. We infer this by analysing the correlation between $\sigma_{z}$ and tracers of each physical process. However, due to the limited time, mass and force resolution of \eagle, we cannot fully isolate the impact of each physical driver. Cosmological simulations with higher mass and force resolution are necessary to understand better the connection between gas turbulence and unresolved processes, such as stellar feedback and gravitational instabilities.

Our results highlight the intricate interplay of different physical processes that determine the level of turbulence in cold gas discs. However, due to the resolution limitations of the \eagle\ simulation (necessary to study unresolved processes such as stellar feedback and gravitational instabilities), we are unable to establish the causality behind high gas turbulence. Nevertheless, our analysis enables us to gauge the relative importance of each mechanism based on halo mass and the epoch at which a galaxy is identified. We infer this by analysing the correlation between $\sigma_{z}$ and tracers of each physical process. Notably, the consistently high correlations found across all halo masses and redshifts for both the angle of gas accretion and the relative amount of cold gas accreted suggest that these mechanisms must be incorporated in explicit models that seek to unravel the primary causes of gas turbulence in discs.

\section*{Data availability}

Observational data for the velocity dispersion will be shared on reasonable request to the corresponding author. Stellar masses and SFRs for the {\sf SAMI}, {\sf DYNAMO}, {\sf KROSS} and {\sc SINS+zC-SINF} surveys are publicly available. The \eagle{} simulations are publicly available; see \citet{McAlpine16, EAGLE17} for how to access \eagle{} data. 

\section*{Acknowledgements}
We thank the anonymous referee for the insightful comments that helped improve the quality of this paper. We thank Hannah \"{U}bler and Nastascha F\"{o}rster Schreiber for providing us with data compilation of ionised velocity dispersion from different surveys. EJ acknowledges the support of the University of Western Australia (UWA) through a scholarship for international research fees and a university postgraduate award. EJ, CL and EW have received funding from the ARC Centre of Excellence for All Sky Astrophysics in 3 Dimensions (ASTRO 3D), through project number CE170100013. CL and ADL are the receipts of an Australian Research Council Discovery Project (DP210101945) funded by the Australian Government. ADL acknowledges financial support from the Australian Research Council through their Future Fellowship scheme (project Nr. FT160100250). This work made use of the supercomputer OzSTAR which is managed through the Centre for Astrophysics and Supercomputing at Swinburne University of Technology. This supercomputing facility is supported by Astronomy Australia Limited and the Australian Commonwealth Government through the national Collaborative Research Infrastructure Strategy (NCRIS). We acknowledge the Virgo Consortium for making their simulation data available. The \eagle\ simulations were performed using the DiRAC-2 facility at Durham, managed by the ICC, and the PRACE facility Curie based in France at TGCC, CEA, Bruyeres-le-Chatel.

%%%%%%%%%%%%%%%%%%%%%%%%%%%%%%%%%%%%%%%%%%%%%%%%%%

%%%%%%%%%%%%%%%%%%%% REFERENCES %%%%%%%%%%%%%%%%%%

\bibliography{Biblio} 
\bibliographystyle{mnras}
% ==============================================================
%                      APPENDIX 
% =============================================================

\appendix

\section{Strong and weak convergence tests} \label{app: Convergence}

In this section, we test whether our results are sensitive to the mass or force resolution used for the intermediate-resolution reference \eagle\ simulation. As in \citet{Schaye15}, we use ``strong convergence'' when comparing runs that differ in mass or force resolution, but not in subgrid physics, and ``weak convergence'' when comparing runs of different resolution, but after recalibrated the subgrid physics of one of them to ensure a similar $z\approx 0$ galaxy population. We carry out our convergence tests using the same cosmological volumes and initial conditions in order to better isolate resolution effects. Table~\ref{tab: Eagle high res} shows the relevant numerical parameters adopted for the \eagle\ reference runs with low ({\sc Ref-L25N376}; corresponding to the mass and force resolution employed for the runs used in our analysis) and high ({\sc Ref-L25N752}) resolution. 

The two runs listed in Table~\ref{tab: Eagle high res} adopt the same subgrid physics and therefore can be used to for strong convergence tests. The \eagle~ suite also includes a ``recalibrated'' run (hereafter {\sc Recal-L25N752}), which used the same softening length and particle masses as the {\sc Ref-L25N752}, but adjusted the subgrid physics parameters to improve the agreement with low-redshift observations of the galaxy population (specifically, the stellar mass function and galaxy size-mass relation). 

%In the latter run, stellar feedback energy per unit stellar mass on the gas density is somewhat different than in the reference runs. However, \citet{Schaye15} showed that $f_{\rm th}$ in equation~(\ref{eq: N-heated}) is almost identical between the {\sc Ref-L25N376} and {\sc Recal-L25N752} runs. The latter also has stronger AGN feedback compared to the {\sc Ref-L25N376} run (see Table~$3$ in \citealt{Schaye15}).

\begin{table}
    \centering
    \begin{tabular}{l|cccccc}
        \hline 
        Parameter & {\sc Ref-L25N376} & {\sc Ref-L25N752}\\
        \hline
        $L$ [cMpc] & 25 & 25\\
        $N_{\rm part}$ & $2\times 376^3$  & $2\times 752^3$ \\
        $m_{\rm gas}$ $\rm [M_{\odot}]$ &  $1.81\times 10^6$ &  $2.26\times 10^5$ \\
        $m_{\rm DM}$ $\rm [M_{\odot}]$ & $9.70\times 10^6$ & $1.21\times 10^6$ \\
       $\epsilon_{\rm com}$ $\rm [ckpc]$& 2.66 & 1.33\\
       $\epsilon_{\rm prop}$ $\rm [pkpc]$& 0.70 & 0.35\\
       \hline
    \end{tabular}
    \caption{Same as Table~\ref{tab: eagle-runs} but for the \eagle\ runs used in this appendix. The {\sc Recal-L25N752} run used the same numerical parameters as {\sc Ref-L25N752}.}
    \label{tab: Eagle high res}
\end{table}

\begin{figure}
	\centering
	\includegraphics[width=\columnwidth]{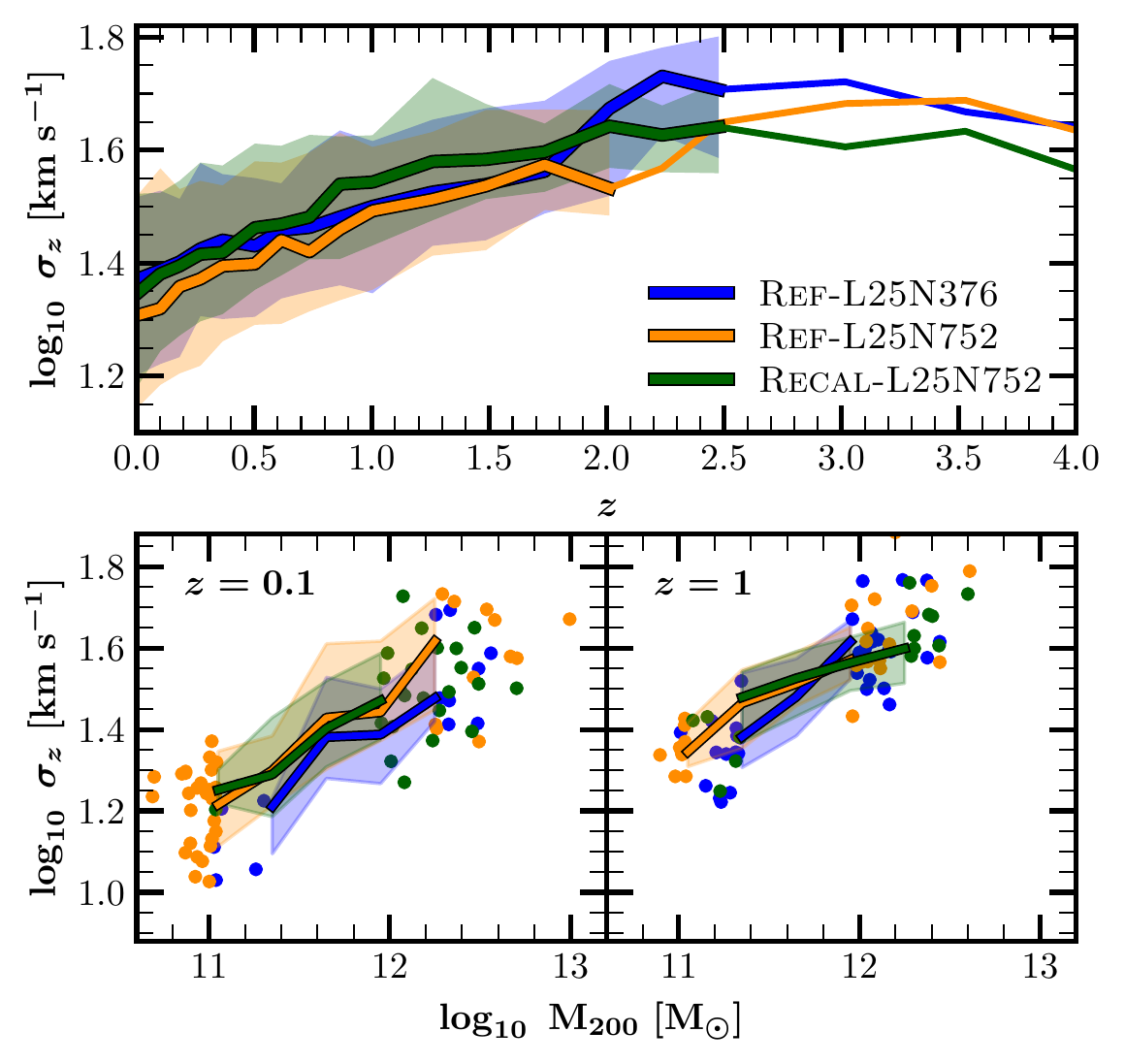}
	\caption{Top panel: The evolution of the gas velocity dispersion for galaxies in the {\sc Ref-L25N376} (blue), {\sc Ref-L25N752} (orange), and {\sc Recal-L25N752} (green) simulations. Thick lines and shaded regions show the medians and the $16^{\rm th}-84^{\rm th}$ percentiles, respectively. Thin lines show the median $\sigma_{z}$ of galaxies in bins containing fewer than $10$ galaxies. Bottom panels: The $\sigma_{z}-\rm M_{200}$ relations for each simulation at $z=0.1$ (left) and $z=1$ (right). Thick lines and shaded regions are as in the top panel. Circles show individual galaxies in bins containing fewer than $10$ objects.}	\label{fig: Convergence}
\end{figure}

Fig.~\ref{fig: Convergence} shows the redshift evolution of $\sigma_{z}$ (top panel) as well as the $\sigma_{z}-\rm M_{200}$ relations at $z=0.1$ (bottom-left) and $z=1$ (bottom-right) for the three simulations mentioned above. Despite the low numbers of galaxies (due to the smaller simulated volume), all three runs exhibit a similar $\sigma_{z}$-redshift relations and $\sigma_{z}- \rm M_{200}$ relations. Moreover, the similarities between results from {\sc Ref-L25N376} and {\sc Ref-L25N752} indicate that the values $\sigma_{z}$ used in our analysis are not particularly sensitive to the resolution adopted for the \eagle~ reference run (i.e. strong convergence is achieved). 

We refer to \citet{Ludlow2020} for a more detailed discussion of strong numerical convergence in the \eagle~ simulation suite.

% =====================================================
\section{Quantifying gravitational instabilities}\label{AppComputingQ}

To investigate the impact of local disc instabilities on gas turbulence we use the framework of Toomre-$Q$ instability. In Section~\ref{ComputingQ} we describe the method to compute the Toomre-$Q$ parameter for galaxies in the {\sc Ref-L100} run. In Section~\ref{AppTestingQ} we test our method on idealised, equilibrium disc galaxy models.  

\subsection{Calculating the Toomre-Q parameter in \eagle} \label{ComputingQ}

We compute a two-component parameter $Q_{\rm net}$ for all galaxies in the selected samples in the {\sc Ref-L100} run. Similar to the $\sigma_z$ calculation (see Section~\ref{sec: sigma_z calculation}), we only include cold gas and star particles enclosed by a cylinder with a height of 6 kpc above and below the disc plane. Our method of computing $Q_{\rm net}$ uses the individual Toomre-$Q$ parameters for stars and gas given by,
\begin{equation} \label{eq: Toomre-Q2}
 Q_i = \frac{\kappa\, \sigma_{r,i}}{\pi\, G \,\Sigma_i},
\end{equation}
\noindent where $i$ is the corresponding baryonic component (stars or gas), $ \kappa$ is the epicyclic frequency, $\sigma_{r,i}$ is the local velocity dispersion, and $\Sigma_i$ is the surface density. Note that for the stellar component, $\pi$ is usually replaced by 3.36. The series of steps to compute each component in equation~(\ref{eq: Toomre-Q2}) are detailed below. \\
 
\noindent {\it Bulge-star removal}: The standard Toomre instability analysis \citep{Toomre64, Binney08} is valid for thin, rotationally supported discs. We therefore remove the bulge component, which deviates from a disc-like morphology. For each star particle, we compute the vertical component of its angular momentum vector, $j_z$, as well as the AM of a circular orbit in the disc plane with the same binding energy, $j_{\rm c}$ (recall that in our analysis the $z$-axis is aligned with the net angular momentum vector of the galaxy disc). We then remove stars with $j_z/j_{\rm c}<0.7$, as these are more likely to be associated with a dispersion supported structure.\\

\noindent {\it The epicyclic frequency:} We calculate the epicyclic frequency, $\kappa$, using  
\begin{equation}
    \kappa^2 = 2\frac{V_{\rm c}^2}{R^2}\left(1 + \beta\right),
\end{equation} \label{eq: epicyclic}
where $V_{\rm c}(R)=(G M /r)^{1/2}$ is the circular velocity at radius $R$ and $\beta = {\rm d}\log V_{\rm c}/{\rm d}\log R$ its logarithmic derivative. We use $R$ to denote the radial distance from the rotation axis. Note that $V_{\rm c}$ is computed using stars (including those in the bulge), gas and DM. \\

\noindent {\it Local velocity dispersions}: We calculate a ``particle-based'' radial and vertical velocity dispersion ($\sigma_r$ and $\sigma_z$, respectively) for gas as follows. For each gas particle, we select its $50$ closest gas neighbours and use them to compute the radial and vertical velocity dispersion. Note that we use the bulk motion of these 50 particles to calculate $\sigma_{r}$ and $\sigma_{z}$. We assign the distance to the $50^{\rm th}$ closest neighbour as the smoothing length of the corresponding particle. Note that this implies that smoothing lengths will be smaller in high-density regions (i.e. central regions in galaxies) and larger for lower densities (i.e. outer disc). We follow the same procedure (using star particles only) to obtain the particle-based $\sigma_r$ and $\sigma_z$ for the stellar component. \\

\noindent {\it Two-dimensional maps using {\sc Py-SPHViewer}}: We grid the face-on projection of discs using square pixels of side $70\ \rm pc$. Using the particle-based $\sigma_{r}$ and the smoothing lengths defined above, we apply the {\sc Py-SPHViewer} code \citep{sphviewer} on the pixelated grid. This produces two-dimensional maps of the surface density and radial velocity dispersion for stars and gas.  We finally use this information to compute the $Q_{\rm gas}$ and $Q_\star$ maps using equation~(\ref{eq: Toomre-Q2}). \\

\noindent {\it Disc thickness correction and two-component $Q_{\rm net}$ parameter}: In thick discs, gravitational forces are weaker, and as a consequence, the $Q_{\rm crit}$ threshold to define  instability can drop below unity \citep[e.g.][]{Goldreich65, Behrendt15}. As we consider \eagle\ discs with a maximum height of $6$~pkpc we must apply a thickness correction to the $Q$ parameters. We apply the correction suggested by \citet{Romeo13} which depends on the ratio of the vertical to radial velocity dispersions. The final effect is to increase the stability parameter, $Q$, by a factor $T$ given by 

\[
  T = 
  \begin{cases}
     1 + 0.6\left(\frac{\sigma_z}{\sigma_r}\right)^2 & \text{if $\left(\frac{\sigma_z}{\sigma_r}\right)<0.5$}, \\
     0.8+0.7\left(\frac{\sigma_z}{\sigma_r}\right) & \text{if $\left(\frac{\sigma_z}{\sigma_r}\right)>0.5$}.\\
  \end{cases} 
  \label{eq: Thickness-correction}
\]

\noindent Hence $Q_{{\rm thick},i} = T\,Q_{\rm i}$, where $i=\{\rm gas, stars\}$. Finally, we calculate the two-component $Q$ parameter that uses the thickness-corrected $Q_{\rm gas}$ and $Q_\star$ using the formulation of \citet{Romeo11}:

\[
  \frac{1}{Q_{\rm net}} = 
  \begin{cases}
     \frac{W}{Q_\star} + \frac{1}{Q_{\rm gas}} & \text{if $Q_{\star}>Q_{\rm gas}$}, \\
     \frac{1}{Q_{\star}} + \frac{W}{Q_{\rm gas}} & \text{if $Q_{\star}<Q_{\rm gas}$}, \\
  \end{cases} 
  \label{eq: multi-Q}
\]

\noindent where 

\[ W = \frac{2\sigma_{\rm stars}\sigma_{\rm gas}}{\sigma_{\rm stars}^2 + \sigma_{\rm gas}^2}. \] 

\noindent This is an improved version of the $Q_{\rm net}^{-1} \approx Q_\star^{-1} + Q_{\rm gas}^{-1}$ approximation of \citet{Wang94}. These $Q_{\rm net}$ maps (see right-panel of Fig.~\ref{fig: Qmaps} for an example) are the ones we use to study the connection between gravitational instabilities and gas velocity dispersion.

% ========================================================
\subsection{Testing the method on idealised galaxies}\label{AppTestingQ}

We test our method on idealised stellar discs embedded in DM haloes. These are drawn from the initial conditions created by {\sc GalIC} \citep{Yurin14}, which are solutions of the collisionless Boltzmann's equation. The disc/halo system is characterised by several parameters and functions, which are the inputs of {\sc GalICs}. These include structural properties of the halo, such as the concentration and virial circular velocity, $V_{200}$, and disc properties, such as its specific angular momentum or the radial and vertical disc structure. Moreover, it is possible to fix the disc's stellar mass fraction, $f_{\rm disc}$ or the mass of the DM particle, $m_{\rm DM}$. 

Here, we limit our analysis to haloes with $V_{200} = 200\ \rm km\ s^{-1}$ sampled with DM particles of $m_{\rm DM} = 10^{7.5}\ \rm M_{\odot}$. Note that this is larger than the DM mass used in the {\sc Ref-L100} run (see Table~\ref{tab: eagle-runs}). We analyse discs with $f_{\rm disc} = \{0.01, 0.03\}$. 

\begin{figure}
	\centering
	\includegraphics[width=\columnwidth]{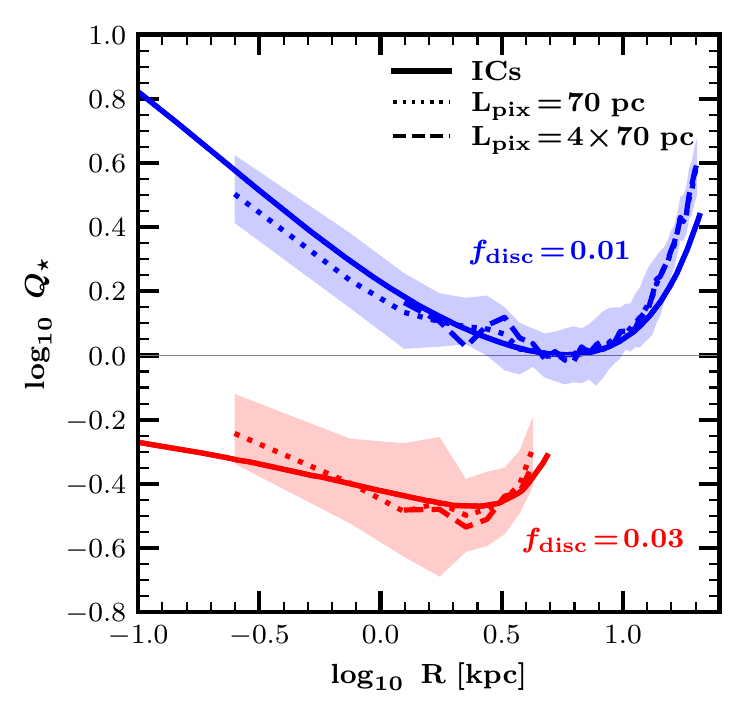}
	\caption{The $Q_{\star}$ profiles of idealised discs as created by {\sc GalIC} (solid), and as calculated by our method (see Section~\ref{ComputingQ}) using pixels of size 70 pc (dotted) and $4\times 70$ pc (dashed). The stellar mass fractions of the discs are $0.01$ (blue) and $0.03$ (red). Lines and shaded regions (the latter are only shown for the dotted lines) correspond to the median and $16^{\rm th}-84^{\rm th}$ percentiles.}
	\label{fig: Q-profiles}
\end{figure}

For $f_{\rm disc}=0.01$, {\sc GalIC} produces a disc that is stable against local perturbations (i.e. $Q_{\star}>1$ across all the disc). Conversely, for $f_{\rm disc}=0.03$, the disc is unstable everywhere. This can be seen in Fig.~\ref{fig: Q-profiles}, where solid lines show the $Q_{\star}$ profiles (from the ICs) for the disc with low (blue) and high (red) stellar mass fractions. To assess whether the $Q_{\star}$ profiles are sensitive to the resolution of the (face-on) 2D maps, we compute the $Q_\star$ radial profiles using $70\ \rm pc$ (dashed lines) and $4\times 70\ \rm pc$ (dotted lines) as the size of the square pixels. We find excellent agreement between the profiles produced by the ICs and those produced by our method. We tested our method using  runs adopting lower and higher $m_{\rm DM}$, and we obtain similar results. 

\label{lastpage}
\end{document}